\documentclass[prd,aps,preprint,showpacs,nofootinbib,eqsecnum,superscriptaddress]{revtex4-1}%twocolumn

\usepackage{hyperref}
\usepackage{amsfonts,amssymb,amsmath,amsthm,amsxtra,mathtools,euscript,mathrsfs}
\usepackage{bm} %-----bold math text               
\usepackage{graphicx}
\usepackage{color,xcolor}
\usepackage{subfigure} %-----use for side-by-side figures
\usepackage{dcolumn} %-----Align table columns on decimal point     
\begin{document}
\title{Equation of State of Neutron Stars with Junction Conditions in the Starobinsky Model}
%---author 1
\author{Wei-Xiang Feng}
\email[Electronic address: ]{wxfeng@gapp.nthu.edu.tw}
\affiliation{Synergetic Innovation Center for Quantum Effects and Applications (SICQEA), 
Hunan Normal University, Changsha 410081, China}
%\affiliation{Chongqing University of Posts \& Telecommunications, Chongqing, 400065, China}
\affiliation{Department of Physics, National Tsing Hua University, Hsinchu 300, Taiwan}

%---author 2
\author{Chao-Qiang Geng}
\email[Electronic address: ]{geng@phys.nthu.edu.tw}
\affiliation{Synergetic Innovation Center for Quantum Effects and Applications (SICQEA), 
Hunan Normal University, Changsha 410081, China}
%\affiliation{Chongqing University of Posts \& Telecommunications, Chongqing, 400065, China}
\affiliation{Department of Physics, National Tsing Hua University, Hsinchu 300, Taiwan}
\affiliation{Physics Division, National Center for Theoretical Sciences, Hsinchu 300, Taiwan}

%---author 3
\author{W.~F.~Kao}
\email[Electronic address: ]{gore@mail.nctu.edu.tw}
\affiliation{Institute of Physics, 
Chiao Tung University, Hsinchu 300, Taiwan}

%---author 4

\author{Ling-Wei Luo}
\email[Electronic address: ]{lwluo@nctu.edu.tw}
\affiliation{Institute of Physics, 
Chiao Tung University, Hsinchu 300, Taiwan}
\affiliation{Department of Physics,
National Tsing Hua University, Hsinchu 300, Taiwan}

%-----------------------------------------------------------------------%
% Abstrct
%-----------------------------------------------------------------------%
\begin{abstract}

We study the Starobinsky or $R^2$ model of $f(R)=R+\alpha R^2$ 
for neutron stars with the structure equations represented by the 
coupled differential equations and the 
\emph{polytropic} type of the matter equation of state. 
The junction conditions of $f(R)$ gravity are used as the boundary 
conditions to match the Schwarschild solution at the surface of the star. 
Based on these the conditions, we demonstrate that 
the coupled differential equations can be solved \emph{directly}. 
In particular, from the dimensionless equation of state
$\bar{\rho} = \bar{k}\, \bar{p}^{\,\gamma}$ with
$\bar{k}\sim5.0$ and $\gamma\sim0.75$ and the constraint of 
$\alpha\lesssim {1.47722}\times 10^{7}\, \text{m}^2$, we obtain the \emph{minimal} mass of 
the NS to be around 1.44 $M_{\odot}$. In addition,
if $\bar{k}$ is larger than 5.0, the mass and radius of the NS would be smaller.

\end{abstract}

%\keywords{}
\date{\today}

%\pacs{}
\maketitle

%-----------------------------------------------------------------------%
% Sec. I - Introduction
%-----------------------------------------------------------------------%
\begin{section}{Introduction}

The astrophysical observations from the
Type Ia Supernovae~\cite{Riess:1998cb,Perlmutter:1998np}, 
large scale structure~\cite{Tegmark:2003ud,Seljak:2004xh} and 
baryon acoustic oscillations~\cite{Eisenstein:2005su} as well as 
cosmic microwave background~\cite{Spergel:2003cb,Spergel:2006hy,Komatsu:2008hk}
indicate the necessity of new physics 
beyond the Einstein's general relativity (GR).
The modified theories of gravity~\cite{Nojiri:2006ri,Nojiri:2017ncd} become more significant 
in order to explain the accelerated expansion phenomenon not only 
inflation~\cite{Starobinsky:1980te,Guth:1980zm,Linde:1981mu,Albrecht:1982wi,Starobinsky:1987zz} 
in the early epoch but also dark 
energy~\cite{Weinberg:1988cp,Sahni:1999gb,Carroll:2000fy,Peebles:2002gy,Padmanabhan:2002ji,Copeland:2006wr}
in the recent stage of the universe. 
A class of alternative theories of the modification from
the \emph{geometric} point of view is the so-called $f(R)$ gravity 
theories~\cite{Barrow:1983rx,Sotiriou:2008rp,DeFelice:2010aj,Nojiri:2010wj}. 
In these theories, the Lagrangian density is modified
by using an arbitrary function $f(R)$ instead of the scalar curvature $R$ of the 
\emph{Einstein-Hilbert} term.
The most well-known $f(R)$ model is the \emph{Starobinsky} or $R^2$ model
with $f(R)=R+\alpha R^2$, originally proposed to obtain the quasi-de Sitter 
solution for inflation~\cite{Starobinsky:1987zz}.
Furthermore, several viable $f(R)$ gravity theories~\cite{Hu:2007nk,Starobinsky:2007hu,Nojiri:2007as,Tsujikawa:2007xu,Cognola:2007zu,ExpGra}
have been used to explain the cosmic acceleration problems.

%--------------------%
In order to realize the structure of a compact star, one needs to know its equation of state (EoS), which characterizes the thermodynamic relation between the density $\rho$, pressure $p$ and temperature $T$ of the dense matter. Under the adiabatic assumption, the EoS is reduced to a \emph{polytropic} relation $\rho=k\,p^{\,\gamma}$. This assumption has been discussed for the neutron stars (NSs) in the literature~\cite{Lattimer:2012nd,Chen:2015zpa,Gandolfi:2015jma,Reisenegger:2015crq,Hebeler:2010jx,Barausse:2007pn,Santos:2009nu,Cooney:2009rr,Babichev:2009fi,Reijonen:2009hi,Orellana:2013gn,Alavirad:2013paa,Ganguly:2013taa,Yazadjiev:2014cza}.
In particular, the allowed region of the polytopes has been shown in~\cite{Hebeler:2013nza}.

%--------------------%
The compact relativistic star was first studied by 
Chandrasekhar \cite{Chandrasekhar}, who assumed that a white dwarf is supported only by 
the completely degenerate electron gas, and then obtained the so-called 
\emph{Chandrasekhar limit} of a white dwarf with the \emph{maximal} mass of 
1.44 $M_{\odot}$. Subsequently, Oppenheimer and Volkoff~\cite{Oppenheimer:1939ne} 
proposed a limit of $0.7\,M_{\odot}$ of a NS by considering 
a completely degenerate neutron gas. However, this approach is inappropriate
due to the strong nuclear repulsive forces of neutrons and 
other \emph{strong interaction} of the heavy hadrons in dense matter.

%--------------------%
In the scenario of GR, the structure of the relativistic stars is determined by 
EoS of matter inside the stars without 
an explicit constraint, whereas
it is expected that the $f(R)$ theories do provide some constraints
with singularity problems~\cite{Briscese:2006xu,Nojiri:2008fk,Frolov:2008uf}. 
The relativistic stars in the modified gravities have been 
studied in the literature~\cite{Kobayashi:2008tq,Babichev:2009td,Upadhye:2009kt,
Santos:2009nu,Cooney:2009rr,Babichev:2009fi,Reijonen:2009hi,
Arapoglu:2010rz,Jaime:2010kn,Santos:2011ye,Orellana:2013gn,
Cheoun:2013tsa,Alavirad:2013paa,Astashenok:2013vza,
Ganguly:2013taa,Yazadjiev:2014cza,Staykov:2014mwa,
Yazadjiev:2015zia,Goswami:2015dma,
Staykov:2015mma,Capozziello:2015yza,Yazadjiev:2015xsj,
Hendi:2015pua,Hendi:2015vta,Bordbar:2015wva,Hendi:2017ibm}.
It has been argued that the compact relativistic stars are \emph{difficult} to 
exist due to the curvature scalar $R$ divergence inside the star
in $f(R)$~\cite{Kobayashi:2008tq}. 
However, the realistic EoS in the 
Starobinsky's dark energy model~\cite{Starobinsky:2007hu}
has been constructed in Ref.~\cite{Babichev:2009td}, in which $R$
does not diverge inside the star,
so that the relativistic stars could occur in $f(R)$.
The pure geometric study is formulated in Ref.~\cite{Ganguly:2013taa},
which imposes the \emph{junction conditions} in $f(R)$~\cite{Senovilla:2013vra,Deruelle:2007pt}
as the additional conditions to solve the \emph{coupled structure equations} and obtain
the final result \emph{indirectly}.

%--------------------%
In this study, we consider the $R^2$ model by performing the calculation 
only in the \emph{Jordan} frame. In our discussions, we solve 
the \emph{coupled structure equations} by the
\emph{junction conditions} approach \emph{directly} rather than 
the perturbation methods~\cite{Cooney:2009rr,Arapoglu:2010rz,Orellana:2013gn,Cheoun:2013tsa}\footnote{Some
other non-perturbative methods have been addressed in Refs.~\cite{Astashenok:2014dja,Capozziello:2011nr,Astashenok:2017dpo}}.
We show that the NSs can exist in the $R^2$ model 
under the polytrope assumption of EoS. 
The \emph{possible} dimensionless EoS 
$\bar{\rho} \sim 5.0\, \bar{p}^{\,0.75}$ is concluded 
by the analysis of the various values of the dimensionless parameter 
$\bar{\alpha}$ in the $R^2$ model,
where the bars represent the dimensionless quantities.
The theoretical constraint on the coefficient $\alpha$ of the $R^2$ term in the model is given by
$\alpha\lesssim 1.47722 \times 10^{7}\, \text{m}^2$. 
By applying the resultant EoS and critical value of $\alpha$, 
the \emph{minimal} mass of the NSs is obtained about 1.44 $M_{\odot}$ 
which is the same as the \emph{Chandrasekhar limit}
of the white dwarf \cite{Chandrasekhar}.
For a fixed parameter $\alpha$,
we observe that the mass and the radius get larger 
when $k$ decreases, while the maximal value of 
$\bar{k}=5.0$ can be illustrated.

%--------------------%
This paper is organized as follows. In Sec.~\ref{sec:SSS of R-squared gravity}, 
we derive the coupled differential equations  and 
show the boundary conditions for the spherically symmetric compact stars
in the $R^2$ model.
In Sec.~\ref{sec:Analysis and Results}, 
we analyze the model parameter $\alpha$ and explore
its reasonable value from the typical units in the neutron star system. 
We discuss our result of EoS under the 
specific choice of the initial conditions. Finally, 
we give conclusions in Sec.~\ref{sec:Conclusions}.

\end{section}

%-----------------------------------------------------------------------%
% Sec. II -  Spherically Symmetric Solution of the $R$-squared model 
%-----------------------------------------------------------------------%
\begin{section}{Spherically Symmetric Solution of the $R^2$ model}
\label{sec:SSS of R-squared gravity}
%---------- Action of the f(R)=>EOM
The action of the $f(R)$ theories with matter is given by
\begin{equation}
S=\frac{1}{2\kappa}\int d^{4}x\sqrt{-g}\,f(R)+S_{m}\,,
\end{equation}
with $\kappa = 8\,\pi$ and the conventional units of $G=c=1$.
By the variation with respect to the metric $g_{\mu\nu}$, 
we have the modified Einstein equations
%---------- modified Einstein equations
\begin{equation}\label{E:Field eq}
f'R^{\mu\nu}-\frac{1}{2}\,fg^{\mu\nu}-
(\nabla^{\mu}\nabla^{\nu} - g^{\mu\nu}\,\square)f'
=\kappa\,T^{\mu\nu}\,,
\end{equation}
with $T^{\mu\nu}$ the energy-momentum tensor and 
$\square=g^{\mu\nu}\nabla_{\mu}\nabla_{\nu}$ the D'Alembertian operator. 
In addition, ``$\,'\,$" in this paper denotes the differentiation with respect to its argument, e.g. $f'(R)=df(R)/dR$.
We will focus on the Starobinsky or $R^2$ model with 
the function of the Lagrangian density
%---------- R-squared model
\begin{equation}\label{E:R-squared}
f(R)=R+\alpha R^2\,.
\end{equation}
%Applying (\ref{E:R-squared}) into (\ref{E:Field eq}),
As a result, we obtain the following field equation
\begin{subequations}\label{E:Rsq}
%---------- tensor eq
\begin{align}\label{E:tensor eq}
  G_{\mu\nu}(1 + 2\,\alpha R) +
  \frac{\alpha}{2}\,g_{\mu\nu}R^2 -
  2\,\alpha\,(\nabla_\mu\nabla_\nu - g_{\mu\nu}\square)R =
  \kappa\,T_{\mu\nu}\,
\end{align}
with $G_{\mu\nu} = R_{\mu\nu} - (1/2)\,R\,g_{\mu\nu}$ the Einstein tensor.
Consequently, the trace equation reads
%---------- trace eq.
\begin{align}\label{E:trace eq}
  -R + 6\,\alpha\,\square R =\kappa\,T\, .
\end{align}
\end{subequations}

%--------------------%
%---------- Ansatz of the metric
In order to study the system of a compact star, 
we will study the solution with an ansatz given by 
the \emph{static} spherical symmetric metric
\begin{align}\label{E:metric}
&ds^2 = -\,e^{2\Phi(r)}\,dt^2+e^{2\Lambda(r)\,}dr^2+r^2\,d\Omega^2\,,
\end{align}
where $\quad d\Omega^2=d\theta^2+\sin^2\theta\,d\varphi^2$
and $\exp(2\Lambda(r)) = (1-2m(r)/r)^{-1}$
with $m(r)$ the \textit{mass function} characterizing the mass enclosed within the radius $r$.
In GR, $m(r)=\int^r_0 4\pi \bar{r}^2\,\rho(\bar{r})\,d\bar{r}$ 
with $\rho(\bar{r})$ the density function. For the radius of the star $r_s$, 
$m(r_s)=M$ can be identified as the \textit{total mass} 
in the \textit{Newtonian limit}. In the $R^2$ model, 
the mass function should be modified with some correction terms.
However, it cannot be integrated by the density function $\rho$ directly.
The function $\Phi(r)$ can be regarded as the 
effective relativistic gravitational potential.
Subsequently, we can obtain the Einstein tensor 
from (\ref{E:metric}), given by
% Einstein tensor
\begin{subequations}
\begin{align}
%---------- tt-component
\label{E:tt-component}
G_{tt}           &= -\frac{1}{r^2}\,e^{2\Phi}\frac{d}{dr}
                    \bigg(r(e^{-2\Lambda} - 1)\bigg)
                  = \frac{2}{r^2}\,e^{2\Phi}\,m{'}\,,\\ 
%---------- rr-component
\label{E:rr-component}
G_{rr}           &= -\frac{1}{r^2}\,e^{2\Lambda}
                    (1-e^{-2\Lambda}) + 
                    \frac{2}{r}\,\Phi{'}\,,\\
%---------- \theta\theta-component
G_{\theta\theta} &= r^2\bigg(\Phi{''} + \Phi{'}^2 - 
                    \Phi{'}\,\Lambda{'} + 
                    \frac{1}{r}\,(\Phi{'} - 
                    \Lambda{'})\bigg)e^{-2\Lambda}\,,\\
%---------- \phi\phi-component
G_{\varphi\varphi}     &= \sin^2 \theta\,G_{\theta\theta}\, .
\end{align}
\end{subequations}
%where the \emph{prime} denotes the derivative with respect to $r$.

%----------------------------------------%
% Subsec. I - The Coupled Differential Equations
%----------------------------------------%
\begin{subsection}{Coupled Differential Equations}
\label{subsec:The Coupled Differential Equations}
%---------- Conservation law & mass parameter ansatz
Considering a \emph{static} perfect fluid with the 
energy-momentum tensor $T^{\mu\nu}=(\rho+p)u^\mu u^\nu+pg^{\mu\nu}$ with $u^{\mu}$, $\rho$, and $p$ denoting the 4-velocity, 
the \textit{density}, and the \textit{pressure} of the fluid respectively.  
The $\nu=r$ component of the \textit{conservation equation} 
$\nabla_\mu T^{\mu\nu}=0$ gives
%---------- Phi derivative
\begin{align}
\quad\Phi{'}=\frac{-p{'}}{\rho+p}\,\label{E:Phi derivative} .
\end{align}
In addition, we can obtain the identity
%---------- Lambda derivative
\begin{align}
\Lambda{'}=\frac{r\,m'-m}{r(r-2m)} \label{E:Lambda derivative}
\end{align}
via the definition of $\Lambda(r)$.
%---------- Static perfect fluid
% $\Phi{'}$ and $\Lambda{'}$ can be written in terms of 
% the \textit{density} $\rho$ and \textit{pressure} $p$, given by
In the local rest frame, $u_t=-e^{\Phi}$ and $u_i=0$ 
with $i=r, \theta$ and $\varphi$ the spatial coordinates, we have
\begin{equation}
T_{tt}           = \rho\,e^{2\Phi}\,,\quad 
T_{rr}           = p\,e^{2\Lambda}\,,\quad
T_{\theta\theta} = p\,r^2\,,\quad
T_{\varphi\varphi}     = p\,r^2\sin^2\theta\,,
\end{equation}
due to $u_\mu u^\mu=-1$.
%---------- Substitute the metric ansatz & conservation law
In addition, we can obtain the following identities for convenience
\begin{align}
\label{E:D'Alembertian scalar curvature}
\square R          &= e^{-2\Lambda}\bigg(R'' + 
              \bigg(\Phi{'} - \Lambda{'} + 
              \frac{2}{r}\bigg)R'\bigg)\,, \\
\label{E:Covariant twice of the scalar curvature}
\nabla_t\nabla_tR &=-e^{2(\Phi-\Lambda)}\Phi^{'}R'\,,\\
\nabla_r\nabla_r R &= R''-\Lambda{'} R'\,.
\end{align}
%By substituting (\ref{E:Phi derivative}) and 
%(\ref{E:Lambda derivative}) into 
%(\ref{E:D'Alembertian scalar curvature})
%and (\ref{E:Covariant twice of the scalar curvature}), 
Consequently, with the metric given by (\ref{E:metric}) and the energy momentum tensor given by the static perfect fluid, we can write the field equations as a set of differential equations
%--------------------%
%---------- Field equation implemented
%The modified Einstein equations can be further
%expressed as three coupled equations $m'=m'(R,R',p,m)$, $p'=p'(R,R',p,m)$ and 
% $R''=R''(R,R',p,m)$.
% Explicitly, we find
%---------- The three coupled equations 
\begin{subequations}\label{E:Coupled ode1}
\begin{align}
\label{E:m-prime equation} 
m'=&\frac{r^2}{12(1+2\alpha R)}\bigg(32\pi\rho+48\pi p+R(2+3\alpha R)\bigg) \nonumber\\
 &-\frac{\alpha(16\pi pr^3+4m(1+2\alpha R)-\alpha r^3R^2-8\alpha R'r(r-2m))R'}{4(1+2\alpha R)(1+2\alpha R+\alpha rR')}\,, \\
\label{E:p-prime equation} 
p' =&  -\frac{(\rho+p)(16\,\pi pr^3 + 4m(1 + 2\,\alpha R) - 
      \alpha\, r^3 R^2 - 8\,\alpha\,rR'(r-2m))}
      {4\,r(1 + 2\,\alpha R + \alpha\,r R'\,)(r - 2m)}\,, \\
\label{E:R-double prime equation} 
 R''=&-\frac{(8\pi(\rho-3p)-R)r^2+12(r-m)\alpha R'}{6\alpha r(r-2m)}
+\frac{r^2((1+3\alpha R)R+16\pi\rho)R'}{6(r-2m)(1+2\alpha R)}+\frac{2\alpha R'^2}{(1+2\alpha R)}\,.
\end{align}
\end{subequations}
Note that we have written  $m'=m'(R,R',p,m)$, $p'=p'(R,R',p,m)$ and $R''=R''(R,R',p,m)$ as algebraic functionals of  $R,R',p,m$.
Here Eq. (\ref{E:m-prime equation}) is derived from the $tt$-component and the trace equation of the field equation (\ref{E:tensor eq}).
% from $tt$-, $rr$- and trace equation of (\ref{E:Rsq}), respectively.
%from (\ref{E:tensor eq}), 
%(\ref{E:trace eq}), (\ref{E:tt-component}), and (\ref{E:rr-component}).
In addition, Eq.~(\ref{E:p-prime equation}) is derived from the $rr$-component of the field equation (\ref{E:tensor eq}), and is also known as the modified Tolman-Oppenheimer-Volkoff (mTOV) equation \cite{Oppenheimer:1939ne}. Finally, Eq. (\ref{E:R-double prime equation}) is derived from the trace equation (\ref{E:trace eq}).

%--------------------%
For the perfect fluid, we assume the EoS is
\emph{polytrope}, i.e.,
\begin{equation}\label{E:Eos}
\rho = k\,p^\gamma\,.
\end{equation}
%---------- The relation of the dimensionless quantities
In order to simplify the calculations,
we can choose the typical values 
$r_*$, $m_*$, $p_*$, $\rho_*$ and $R_*$
for the compact star system and express  
$r\equiv xr_{*}$, $m\equiv\bar{m}m_{*}$, $p\equiv\bar{p}p_{*}$, 
$\rho\equiv\bar{\rho}\rho_{*}$, $R\equiv\bar{R}R_{*}$ 
and $\alpha\equiv\bar{\alpha}\alpha_*\equiv\bar{\alpha}(1/R_{*})$
in terms of the dimensionless quantities
$x$, $\bar{m}$, $\bar{p}$, $\bar{\rho}$, $\bar{R}$ and $\bar{\alpha}$, 
while the derivatives of $p$, $m$ and $R$ can be written as   
$p'=\bar{p}'(p_{*}/r_{*})$, $m'=\bar{m}'(m_{*}/r_{*})$, $R'=\bar{R}'(R_{*}/r_{*})$ 
and $R''=\bar{R}''(R_{*}/r_{*}^2)$, respectively,
where the \emph{prime} of the dimensionless quantities denotes 
the derivative with respect to $x$.
The polytropic type of EoS in terms of the dimensionless quantities 
can be given as $\bar{\rho}=\bar{k}\,\bar{p}^{\,\gamma}$ with $\bar{k}=k{\rho^{-1}_*} p_*^{\gamma} $.
Since we are interested in the NSs in the $R^2$ model, 
it is convenient for us to define the following typical values in SI units,
%---------- The dimensional quantities 
\begin{align*}
m_{*} &\equiv M_{\odot} = 1.99\times10^{30}\,\text{kg}\,,\\
r_{*} &\equiv 10^{4}\,\text{m} = 10\,\text{km}\,,\\
\rho_{*} &\equiv \frac{\text{Neutron mass}}
         {(\text{Neutron Compton wavelength})^3}
         \sim 10^{18}\,\text{kg}/\text{m}^3\,,\\
p_{*} &=\rho_{*} = 8.99\times10^{34}\,\text{Pa} = 
        8.99\times10^{34}\,
           \text{kg}\,\text{m}^{-1}\,\text{s}^{-2}\,,\\
R_{*} &=\rho_{*} = 7.42\times10^{-10}\, \text{m}^{-2} = 
        7.42\times10^{-4}\, \text{km}^{-2}\,.
\end{align*}
According to the typical units, we can rewrite 
(\ref{E:m-prime equation}), (\ref{E:p-prime equation})
and (\ref{E:R-double prime equation}) as 
the dimensionless equations:
%---------- The dimensionless three coupled equations 
\begin{subequations}\label{E:Coupled ode2} 
\begin{align}
\label{E:Bar m-prime equation}
  \bar{m}{'}=&\frac{x^2}{12(1+2\bar{\alpha}\bar{R} )}\bigg(32\pi\bar{\rho}
                 +48\pi\bar{p}+\bar{R}(2+3\bar{\alpha}\bar{R})\bigg)\bigg(\frac{\rho_{*}r_{*}^3}{m_{*}}\bigg)
                 \nonumber\\&-\frac{\bar{\alpha}((16\pi\bar{p}-\bar{\alpha}\bar{R}^2)
                 x^3(R_{*}r_{*}^{2})+4\bar{m}(1+2\bar{\alpha}\bar{R})
                 (\frac{m_{*}}{r_{*}})-8\bar{\alpha}x(x-2\bar{m}(\frac{m_{*}}{r_{*}}))
                 \bar{R}{'})\bar{R}{'}}{4(1+2\bar{\alpha}\bar{R})(1+2\bar{\alpha}\bar{R}+
                 \bar{\alpha}x\bar{R}')(\frac{m_{*}}{r_{*}})}\,, \\
\label{E:Bar p-prime equation}
\bar{p}{'}  =& -\frac{(\bar{\rho} + \bar{p})(x^{3}(16\pi\bar{p} - 
               \bar{\alpha}\bar{R}^{2})(R_{*}r_{*}^{2}) + 
               4\bar{m}(1 + 2\bar{\alpha}\bar{R})(\frac{m_{*}}
               {r_{*}}) - 8x\bar{\alpha}\bar{R}{'}(x - 2\bar{m}
               (\frac{m_{*}}{r_{*}})) )}
               {4\,x(x - 2\bar{m}(\frac{m_{*}}{r_{*}}))
               (1 + 2\bar{\alpha}\bar{R} + \bar{\alpha}x\bar{R}')}\,, \\
\label{E:Bar R-double prime equation}
\bar{R}{''}=&-\frac{x^2(8\pi(\bar{\rho}-3\bar{p})-\bar{R})(R_{*}r_{*}^{2})
               +12(x-\bar{m}(\frac{m_{*}}{r_{*}}))\bar{\alpha}\bar{R}{'}}{6\bar{\alpha}x
               (x-2\bar{m}(\frac{m_{*}}{r_{*}}))}\nonumber\, \\
               &+\frac{x^2((1+3\bar{\alpha}\bar{R})+16\pi\bar{\rho})\bar{R}'
               (R_{*}r_{*}^{2})}{(x-2\bar{m}(\frac{m_{*}}{r_{*}}))
               (1+2\bar{\alpha}\bar{R})}+\frac{2\bar{\alpha}\bar{R}{'}^2}{1+2\bar{\alpha}\bar{R}}\,,
\end{align}
\end{subequations}
respectively. The dimensionless parameters $m_*/r_*=0.147688$ and $R_{*} r_*^2=0.074215$ characterize the compactness of the NS. In order to discuss the structure of the
NS, we have to solve the three coupled equations 
(\ref{E:Bar m-prime equation}), (\ref{E:Bar p-prime equation}) and 
(\ref{E:Bar R-double prime equation}) numerically with the EoS $\bar{\rho}=\bar{k}\,\bar{p}^{\,\gamma}$.

\end{subsection}

%----------------------------------------%
% Subsec. II - Boundary conditions
%----------------------------------------%
\begin{subsection}{Boundary Conditions}
\label{subsec:Boundary conditions}
%{\color{red}
In GR, the Birkhoff's theorem states that the spherically symmetric vacuum 
solution must be given by the Schwarzschild metric. On the other hand, 
even though the absence of the Birkhoff's theorem in $f(R)$ theories 
might lead to the non-uniqueness of this vacuum solution, 
the Schwarzschild metric
can serve as a vacuum solution in $f(R)$  under some circumstances. 
It has been shown that
the conditions of  $R=0$ with $f(0)=0$ and $f'(0)\neq 0$ for the existence of the Schwarzschild metric
 are satisfied in the Starobinsky model~\cite{Nzioki:2013lca}.
As a result, we introduce the Schwarzschild vacuum solution for the exterior region.
In this way, we can obtain the mass and radius of the star from 
the Schwarzschild metric once (\ref{E:Coupled ode2}) is solved 
with proper boundary conditions.
%}

In the following, we consider the star without thin shells.
In order to match the solution at the {\emph{surface of the star}}, 
we use the Schwarzschild solution 
{for the exterior region ($r>2\tilde{M}$)}
\begin{equation}
ds^2 = -\bigg(1 - \frac{2\tilde{M}}{r}\bigg)\,dt^2 + 
      \bigg(1 - \frac{2\tilde{M}}{r}\bigg)^{-1}\,dr^2 + 
      r^2\,d\Omega^2,
\end{equation}
where $\tilde{M}$ is the mass parameter in GR. 
The \emph{junction conditions} for the $f(R)$ theories should be 
more restrictive as discussed in 
Refs.~\cite{Senovilla:2013vra,Ganguly:2013taa,Deruelle:2007pt}.
The \emph{first} and the \emph{second} fundamental forms of the conditions 
are $[h_{\mu\nu}] = 0$ and $[K_{\mu\nu}] = 0$, respectively, where $[\,]$ denotes 
the jump at the boundary surface of the star.
We can identify $\tilde{M}$ with $M=m(r_{s})$ only when the first fundamental form matches.
However, there are \emph{two} additional conditions
 for the scalar curvature across the surface \cite{Ganguly:2013taa}, given by
%---------- junction conditions 
\begin{subequations}
\begin{align}
[R]             &= 0\,,\\
[\nabla_{\mu}R] &= 0\,.
\end{align}
\end{subequations}
In our assumption with the static and spherically symmetric metric,
the curvature $R$ is only a function of $r$.
%---------- Boundary conditions at the surface
By matching of the second fundamental form to make 
the pressure vanishing at the boundary surface \cite{Senovilla:2013vra},
the boundary conditions are reduced to $R(r_{s})=0$, 
$R'(r_{s})=0$ and $p(r_{s})=0$.
Inside the star, we have to determine the boundary 
conditions at the center of the star. There are two first-order 
and one second-order differential equations in Eq.~(\ref{E:Coupled ode1}). 
Hence, only \emph{four} boundary conditions are required 
to solve these coupled ordinary differential equations.
%---------- Boundary conditions at the center
To satisfy the regularity conditions at 
the center of the star, we must have 
$m(0)=0$, $p'(0)=0$, $\rho{'}(0)=0$ and $R'(0)=0$ \cite{Ganguly:2013taa},
in which two of them are redundant.
According to Eq.~(\ref{E:p-prime equation}), $p'(0)=0$ is 
automatically satisfied as long as $m(0)=0$ and 
$R'(0)=0$ as $r\rightarrow 0$. In addition, 
$\rho$ and $p$ are related by EoS in (\ref{E:Eos}),
leading to $p'(0)=0$ and $\rho{'}(0)=0$,
so that only conditions $m(0)=0$ and $R'(0)=0$ are left.

%--------------------%
Consequently, we have three boundary conditions 
at the surface and two boundary ones at the center
written in the dimensionless forms, given by
%---------- physical requirements
\begin{equation}\label{E:Boundary conditions}
\bar{R}(x_{s})=0\,,\quad \bar{R}{'}(x_{s})=0\,,\quad 
\bar{p}(x_{s})=0\,,\quad \bar{m}(0)=0\,,\quad 
\bar{R}{'}(0)=0\,.
\end{equation}
These boundary conditions are referred to as 
the \emph{Schwarzschild boundary conditions}.
Mathematically, since there are four undetermined integration constants $c_1, c_2, c_3$ and $c_4$ in (\ref{E:Coupled ode2}), only \emph{four} in (\ref{E:Boundary conditions}) are enough to solve it. However, these integration constants should be associated with the model parameter $\alpha$ and $(\gamma, \bar{k})$ in the EoS. The fifth one in (\ref{E:Boundary conditions}) can be used to constrain the parameter space of $(\alpha, \gamma, \bar{k})$.
For example, if we choose $\bar{m}(0)=\bar{R}{'}(0)=\bar{p}(x_{s})=\bar{R}(x_{s})=0$, then we have to determine whether $\bar{R}{'}(\alpha, \gamma, \bar{k}; x)|_{x=x_{s}}$ satisfies $\bar{R}{'}(x_{s})=0$ for fixed values of $\alpha$, $\gamma$ and $\bar{k}$. 

%--------------------%
According to the mTOV equation in (\ref{E:Coupled ode1}) and 
conservation equation in (\ref{E:Phi derivative}), we have
\begin{equation}\label{gradient phi}
\frac{d\Phi}{dr} = \frac{16\pi pr^3+4m(1 + 2\alpha R) 
                   - \alpha r^{3}R^{2} - 8\alpha r(r - 2m)R'}
                   {4r(1+2\alpha R + \alpha rR')(r-2m)}.
\end{equation}
In the region outside of the star ($r\geq r_{s}$), the pressure and 
scalar curvature as well as the derivative of the scalar curvature should be continuous, 
resulting in $p(r)=0$, $R(r)=0$ and $R'(r)=0$ by (\ref{E:Boundary conditions}). 
It can be checked that the exterior solution of (\ref{gradient phi}) 
coincides with the Schwarzschild solution 
$e^{2\Phi(r)}=1-2\tilde{M}/r$.

\end{subsection}
\end{section}

%-----------------------------------------------------------------------%
% Sec. III - Results and Analysis
%-----------------------------------------------------------------------%
\begin{section}{Analysis and Results}
\label{sec:Analysis and Results}
%----------------------------------------%
% Subsec. I - The Determination of $\alpha$
%----------------------------------------%
\begin{subsection}{Determination of $\alpha$}
\label{subsec:The singular problem on alpha}
In principle, Eq.~(\ref{E:Coupled ode1}) can be regarded as 
the GR results with $\alpha R^2$ as the modification  term.
For example, Eq.~(\ref{E:p-prime equation}) corresponds to 
the TOV equation in GR \cite{Oppenheimer:1939ne} when $\alpha\rightarrow 0$.
%---------- Important observations
Similarly, we can recover $m'$ and $R''$ equations in GR for
(\ref{E:m-prime equation}) and (\ref{E:R-double prime equation}) 
with $\alpha\rightarrow 0$.
By separating the GR contribution, Eq.~(\ref{E:m-prime equation}) 
can be rewritten as 
\begin{align} \label{E:m-prime equation 2}
m' = 4\pi r^2\rho&-\frac{r^2(8\pi(\rho-3p)-R)}{6(1+2\alpha R)} - 
     \frac{\alpha Rr^2}{4(1 + 2\alpha R)}(32\pi\rho - R)\, \nonumber\\
  &-\frac{\alpha(16\pi pr^3+4m(1+2\alpha R)-\alpha r^3R^2-8\alpha r(r-2m)R')R'}{4r(1+2\alpha R)(1+2\alpha R+\alpha rR')}\equiv 4\pi r^2\rho_{\text{eff}}\,,
\end{align}
where
\begin{align} 
\rho_{\text{eff}} = \rho&-\frac{8\pi(\rho - 3p) - R}{24\pi(1+2\alpha R)} - 
             \frac{\alpha R}{16\pi(1+2\alpha R)}(32\pi\rho - R)\, \nonumber\\
             &-\frac{\alpha(16\pi pr^3+4m(1+2\alpha R)-\alpha r^3R^2-8\alpha r(r-2m)R')R'}{16\pi r^3(1+2\alpha R)(1+2\alpha R+\alpha rR')} .
\end{align} 
In the limits of $\alpha\rightarrow 0$ and $R \rightarrow 8\pi(\rho-3p)$, 
we have $m'\rightarrow4\pi r^2\rho$, which is the same as result in GR.

However, in the numerical analysis, there are problems of choosing $\alpha$ for the system. 
On one hand, the main numerical difficulty arises from
(\ref{E:R-double prime equation}), in which
\begin{equation}\label{E:R-double prime with p-prime and m-prime}
R''=-\frac{(8\pi(\rho-3p)-R)r^2}{6\alpha r(r-2m)}+\frac{12(r-m)R'}{6r(r-2m)}
+\frac{r^2(R+16\pi\rho)R'}{6(r-2m)}
\end{equation}
by taking  $\alpha\rightarrow 0$.
Furthermore, we have the boundary conditions $R'(0)=0$ and $m(0)=0$ as $r\rightarrow0$,
and obtain
\begin{equation}\label{E:R-double prime limit}
\mathcal{R}''\equiv R''|_{r\rightarrow0}= -\frac{8\pi(\rho-3p)-R}{6\alpha}\,\bigg |_{r\rightarrow0},
\end{equation}
which implies the singularity of $\mathcal{R}''$ as $\alpha\rightarrow 0$ 
under the numerical calculation.
As a result, we encounter the fine-tuning problem of $p(0)$ and $R(0)$. 
On the other hand, we would like to discuss the upper bound for $\bar{\alpha}$.
In the dimensionless form $x=r/r_*$, 
(\ref{E:Bar m-prime equation}) with (\ref{E:m-prime equation 2}) 
and (\ref{E:R-double prime limit}) in $x\rightarrow 0$ 
can be read as 
\begin{equation}\label{E:Bar m-prime equation 2}
\bar{\bm{m}}{'} = x^2\bigg(4\pi\bar{\rho} - \frac{8\pi(\bar{\rho} 
                  - 3\bar{p}) - \bar{R}}{6(1+2\bar{\alpha}\bar{R})} 
                  - \frac{\bar{\alpha}\bar{R}}{4(1 + 2\bar{\alpha}\bar{R})}(32\pi\bar{\rho} 
                  - \bar{R})\bigg)\bigg(\frac{\rho_{*}r_{*}^3}{m_{*}}\bigg)\bigg 
                  |_{x\rightarrow0} \,,
\end{equation}
and
\begin{equation}\label{E:Bar R-double prime dimensionless limit}
\bar{\mathcal{R}}'' \equiv \bar{R}'' |_{x\rightarrow0} 
                    = -\frac{8\pi(\bar{\rho} - 3\bar{p})-\bar{R}}
                    {6\bar{\alpha}}(R_{*}r_{*}^2)\bigg |_{x\rightarrow0}
\end{equation}
respectively, where $(\rho_{*}\,r_{*}^3)/m_{*}={ 0.502513}$
and $R_{*}r_{*}^2=\rho_{*}r_{*}^2 = 7.42\times10^{-2}$,
which characterizes the compactness of a star.
In order to determine the proper value of $\bar{\alpha}$,
we use (\ref{E:Bar R-double prime dimensionless limit}) to rewrite 
(\ref{E:Bar m-prime equation 2}) as
\begin{equation}
\bar{\bm{m}}{'} = 4\pi x^2\bar{\rho}\bigg(\frac{\rho_{*}r_{*}^3}{m_{*}}\bigg) + 
             \bigg(\frac{\bar{\alpha}x^2}{1+2\bar{\alpha}\bar{R}}\bigg)
             \bigg[\bar{R}{''}\bigg(\frac{r_{*}}{m_{*}}\bigg) - 
             \bar{R}\bigg(8\pi\bar{\rho} - 
             \frac{\bar{R}}{4}\bigg)\bigg(\frac{\rho_{*}r_{*}^3}{m_{*}}\bigg)
             \bigg]\bigg |_{x\rightarrow0}\,.
\end{equation}
Since $\bar{R}$ is \emph{convex upward} around $x=0$,  we expect that
$\bar{\mathcal{R}}{''}\leq 0$. Then, we have $8\pi(\bar{\rho}-3\bar{p}) - 
\bar{R}\geq 0$ for $\bar{\alpha}\geq 0$, which can be seen from 
(\ref{E:Bar R-double prime dimensionless limit}).
We can choose 
$\bar{\alpha}\lesssim (m_{*}/r_{*})(R_{*}r_{*}^2)={ 0.010961}$
and obtain the inequality
\begin{equation}
\bar{\bm{m}}{'} \gtrsim 
    4\pi x^2\bar{\rho}\bigg(\frac{\rho_{*}r_{*}^3}{m_{*}}\bigg) + 
    \bigg(\frac{x^2}{1+2\bar{\alpha}\bar{R}}\bigg)
    \bigg[\bar{R}{''}-\bar{R}\bigg(8\pi\bar{\rho} - 
    \frac{\bar{R}}{4}\bigg)(R_{*}r_{*}^2)\bigg]
    (R_{*}r_{*}^2)\bigg |_{x\rightarrow0}\,.
\end{equation}
The last two terms in the square bracket represent  
the first-order and second-order corrections in the $R_{*}r_{*}^{2}$ unit, respectively.
Therefore, we derive
$\alpha=\bar{\alpha}/R_{*}\lesssim 1.47722\times 10^{7}\, \text{m}^2$. 
In addition, several constraints on $\alpha$ from the observational data have been derived in ~\cite{Arapoglu:2010rz,Cheoun:2013tsa,Naf:2010zy}. Moreover, Gravity Probe B~\cite{Everitt:2009} gives $\alpha\lesssim5\times10^{11}\, \text{m}^2$; the precession measurement of the pulsar B in the PSR J0737-3039 system~\cite{Breton:2008xy} yields $\alpha\lesssim2.3\times10^{15}\, \text{m}^2$; and the strong magnetic NS~\cite{Arapoglu:2010rz,Cheoun:2013tsa} results in $\alpha\lesssim10^{5}\, \text{m}^2$.
Furthermore, it has been shown that
the \emph{ghost-free} condition $f''(R)\geq0$ \cite{Sotiriou:2008rp} leads to $\alpha>0$. 
Here, it should be noted that 
only within the condition $\bar{R} \rightarrow 8\pi(\bar{\rho}-3\bar{p})$
can we have a finite $\bar{\mathcal{R}}{''}$ in the limit $\bar{\alpha}\rightarrow 0$. 
This condition assures that the $R^2$ model is consistent with GR in $\alpha\rightarrow 0$.
\end{subsection}

%----------------------------------------%
% Subsec. II - Numerical results
%----------------------------------------%
\begin{subsection}{Numerical Results}
\label{subsec:Numerical results}
%--------------------%
By using the Runge-Kutta 4th-order (RK4) procedure, 
Eq.~(\ref{E:Coupled ode2}) can be solved by choosing $\bar{p}(0)$ 
and $\bar{R}(0)$ as the central values with boundary conditions 
$\bar{m}(0)=\bar{R}{'}(0)=\bar{p}(x_{s})=0$. We can obtain 
$\bar{R}(x_{s})$ and $\bar{R}{'}(x_{s})$ by applying 
random values of $\bar{p}(0)$ and $\bar{R}(0)$ numerically. 
In terms of the problem of (\ref{E:R-double prime limit}), we have to find out the appropriate values of $\bar{p}(0)$ and $\bar{R}(0)$
to satisfy $\bar{R}(x_{s})=0$ and $\bar{R}{'}(x_{s})=0$, 
which maintain the \emph{Schwarzschild boundary conditions}(\ref{E:Boundary conditions}).
%--------------------%
%%%%%%%%%%%%%%%%%%%%%%%%%%%%%%%%%%%%%%%%%%%%%
%Table of Mass, Radius, Ricci curvature at center, Polytropic exponent, Polytropic constant
%%%%%%%%%%%%%%%%%%%%%%%%%%%%%%%%%%%%%%%%%%%%%
\begin{table}
  \caption{The results of the radius $x_{s}=r_{s}/r_{*}$ and mass $\bar{M}$ 
  with the polytropic exponent $\gamma$ and central Ricci curvature $\bar{R}(0)$
  for the $R^2$ model respect to the various $\bar{\alpha}$ with the fixed central pressure $\bar{p}(0)=1$ and polytropic constant $\bar{k}=5.0$.}
 %\centering
 \begin{ruledtabular}
 \begin{tabular}{c c c c c}
  %\hline\hline
  % inserts table 
  %heading
  $\bar{\alpha}$  &  $x_{s}$  &  $\bar{M}$  &  $\gamma$    &  $\bar{R}(0)$ \\ [0.5ex] 
  \hline
  0.01            &  1.999     &  1.444      &  0.7525000000  &  8.95 \\
  0.0005          &  2.477     &  1.557      &  0.7503553926  &  35.00 \\ 
  GR ($\alpha=0$) &  2.297     &  1.672      &  0.7503553926  &  16$\pi$\\[1ex]
  %\hline\hline
 \end{tabular}
 \end{ruledtabular}
 \label{table:I}
\end{table}

%--------------------%
%%%%%%%%%%%%%%%%%%%%%%%%%%%%%%%%%%%%%%%%%%%%%
%Various profiles in the interior of the star for alpha=0.0005,kb=4.5 %%%%%%%%%%%
%%%%%%%%%%%%%%%%%%%%%%%%%%%%%%%%%%%%%%%%%%%%%
\begin{figure}[b]
  %\centering
  \subfigure[]{
    \includegraphics[width=0.5\textwidth]{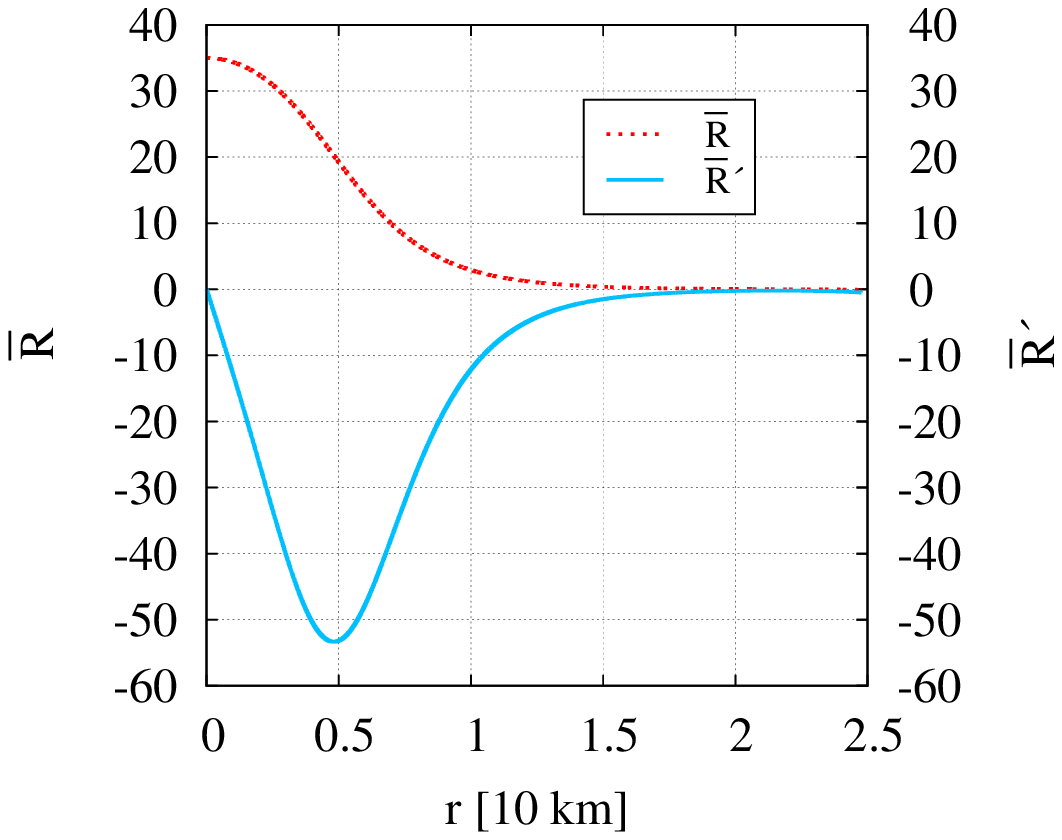}}
  ~
  \subfigure[]{
    \includegraphics[width=0.5\textwidth]{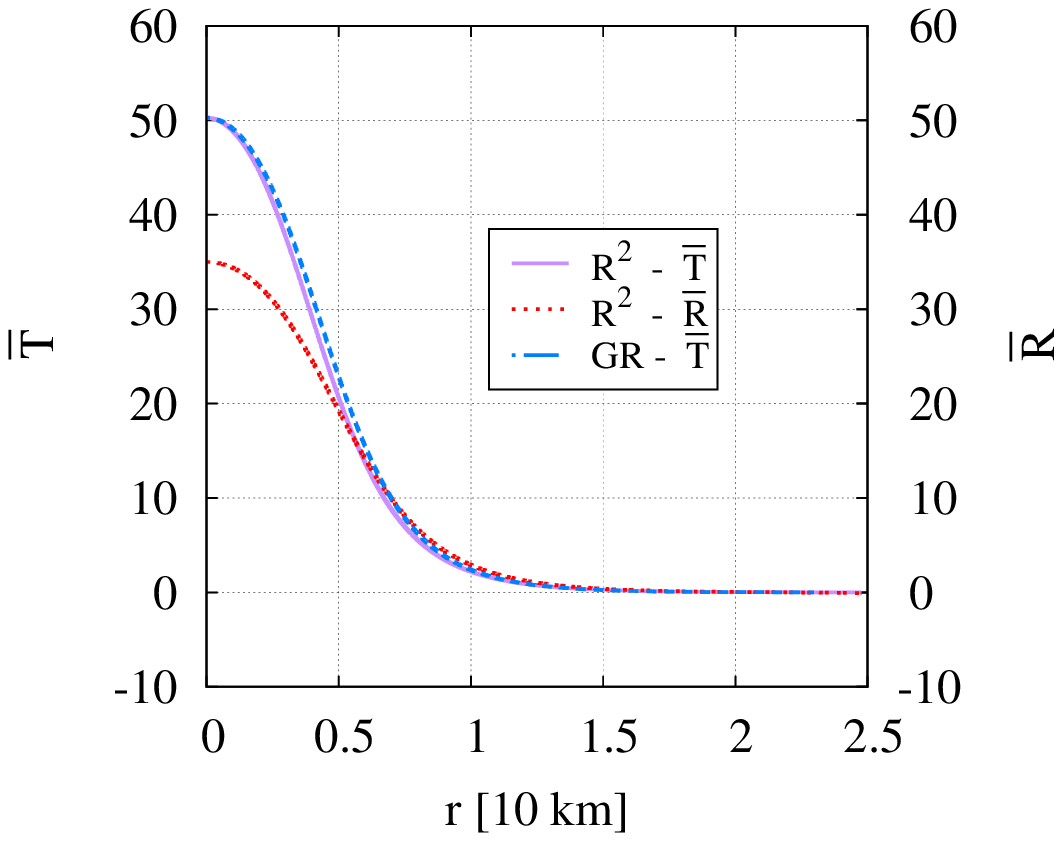}}
  \caption{(color online) (a) The curvature scalar 
  $\bar{R}$ (dotted line) and the derivative of the curvature scalar 
  $\bar{R}'$ (solid line) of the radial coordinate $r$ in the unit of 10 km and 
  (b) the profiles of the curvature $\bar{R}$ of 
  the $R^2$ model (dotted line)
  and the negative trace of the energy momentum tensor 
  $\bar{T}:=8\pi(\bar{\rho}-3\bar{p})$ of the 
  $R^2$ model (solid line) 
  and GR (dotted long-dashed line), where $\bar{k}=5.0$ and $\bar{\alpha}=0.0005$. }
  \label{fig:I}
\end{figure}

%--------------------%
The parameters $\bar{k}$ and $\gamma$ affect the behaviors of the 
coupled equations (\ref{E:Coupled ode2}) as well as the 
boundary values at the surface. Clearly, they can be determined 
once our boundary conditions are fixed in the numerical calculations.
All the results are given in the 
typical units $m_{*}$,\footnote{$m_{*}=M_{\odot}$} 
$r_{*}$,\footnote{$r_{*}=10$ km} $\rho_{*}$, $p_{*}$, 
and $R_{*}$ as defined in Sec.~IIA.
We look for the reasonable EoS for 
$\bar{\alpha}=0.01$ and $0.0005$ and compare the results with GR ($\alpha=0$). 
For simplicity, we keep the high pressure at the center 
of the star to be $\bar{p}(0)=1$ initially. Then, we end up the calculation
with $\bar{p}(x_{s})=10^{-6}$ at the surface of the star, 
corresponding to the density at the \emph{bottom} of the NS's
outer crust around $10^{13}\sim10^{14}\,\text{kg}\,\text{m}^{-3}$.
We keep $\bar{k}=5.0$ and fine-tune the parameters $\gamma$ and $\bar{R}(0)$ 
in order to satisfy $\bar{R}(x_s)=0$ and $\bar{R}'(x_s)=0$.
The results are given in the TABLE \ref{table:I}.
%{\color{red}
From this table, we find that \emph{for a smaller 
$\bar{\alpha}$, $\bar{R}(0)$ is larger}, and the same goes for $\bar{M}$,
which are the generic feature of the  model. The behaviors of
the growing $\bar{\alpha}$ and decreasing $\bar{M}$ have been also 
discussed in Ref.~\cite{Astashenok:2017dpo} with the realistic EoS instead of the polytropic one in this study. 
We note that the different choices of $\bar{k}$ will be shown in  TABLE \ref{table:II}.
%--------------------%
\begin{figure}
  %\centering
  \subfigure[]{
    \includegraphics[width=0.5\textwidth]{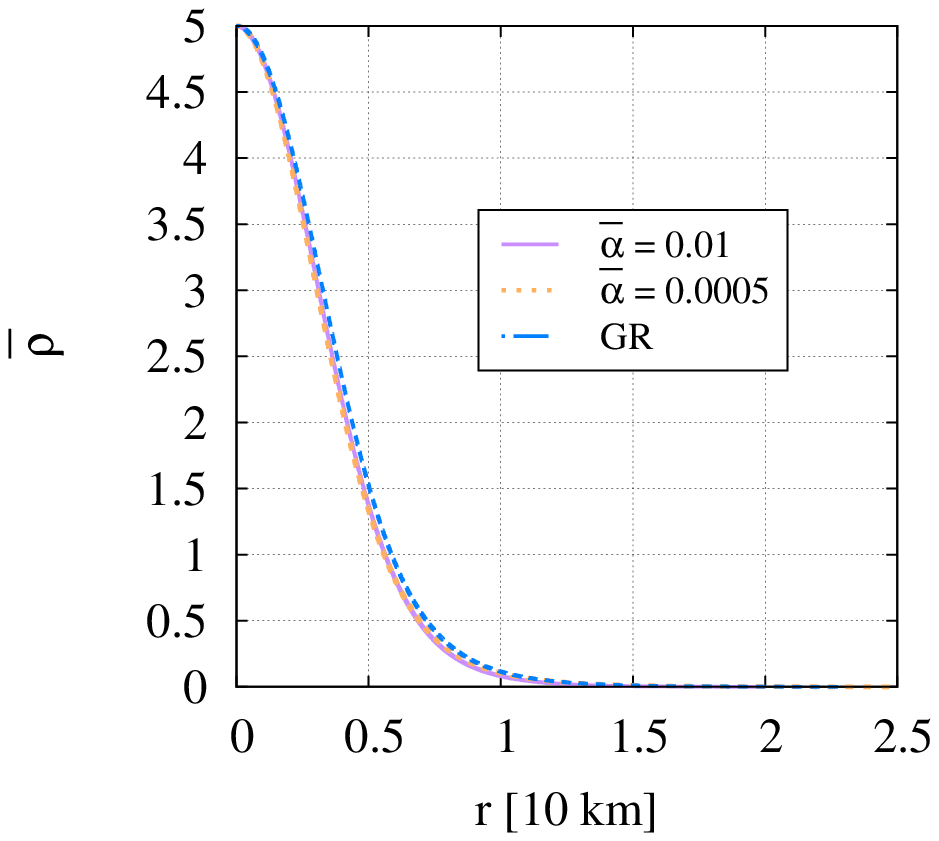}}
  ~
  \subfigure[]{
    \includegraphics[width=0.5\textwidth]{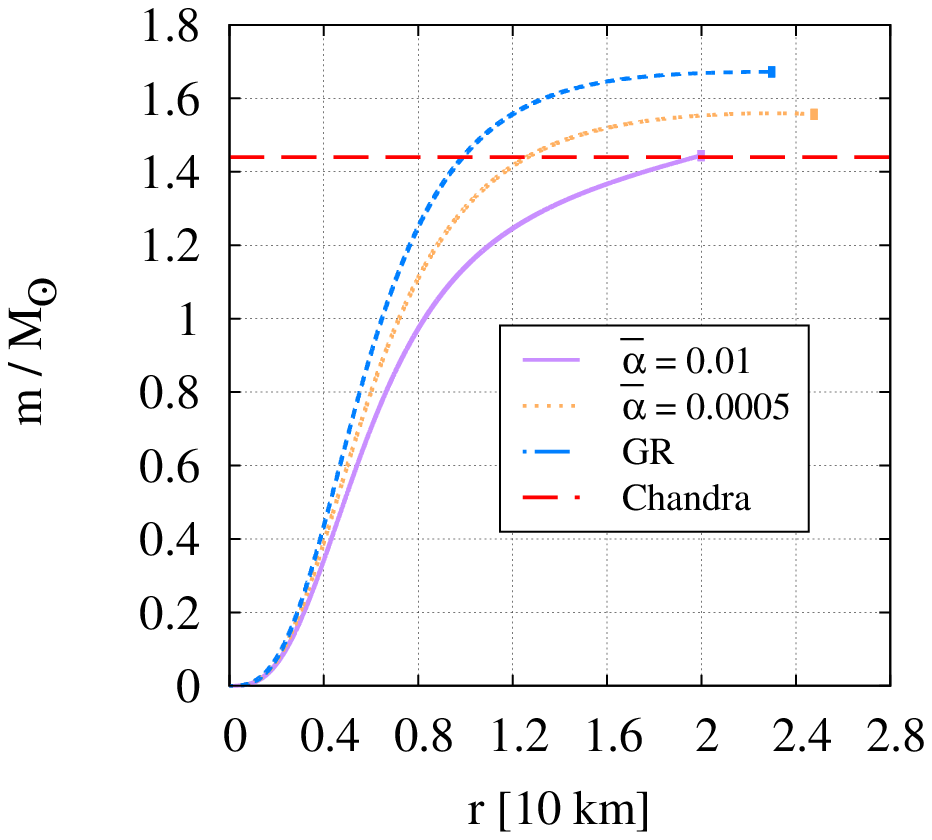}}
  \caption{
  (color online)
  (a) The density $\bar{\rho}$ and
  (b) mass $m$ as functions of the 
  radial coordinate $r$ in the unit of 10 km with $\bar{k}=5.0$, where 
  the solid and dotted lines indicate the $R^2$ model 
  with $\bar{\alpha}=0.01$ and $0.0005$, respectively,
  and the dotted long-dashed line corresponds to the GR case,
  while the value 1.44 is represented as   
  the Chandrasekhar (Chandra) limit 
  (long-dashed line)
  }
  \label{fig:II}
\end{figure}

%--------------------%
\begin{figure}
  %\centering
    \includegraphics[width=\textwidth]{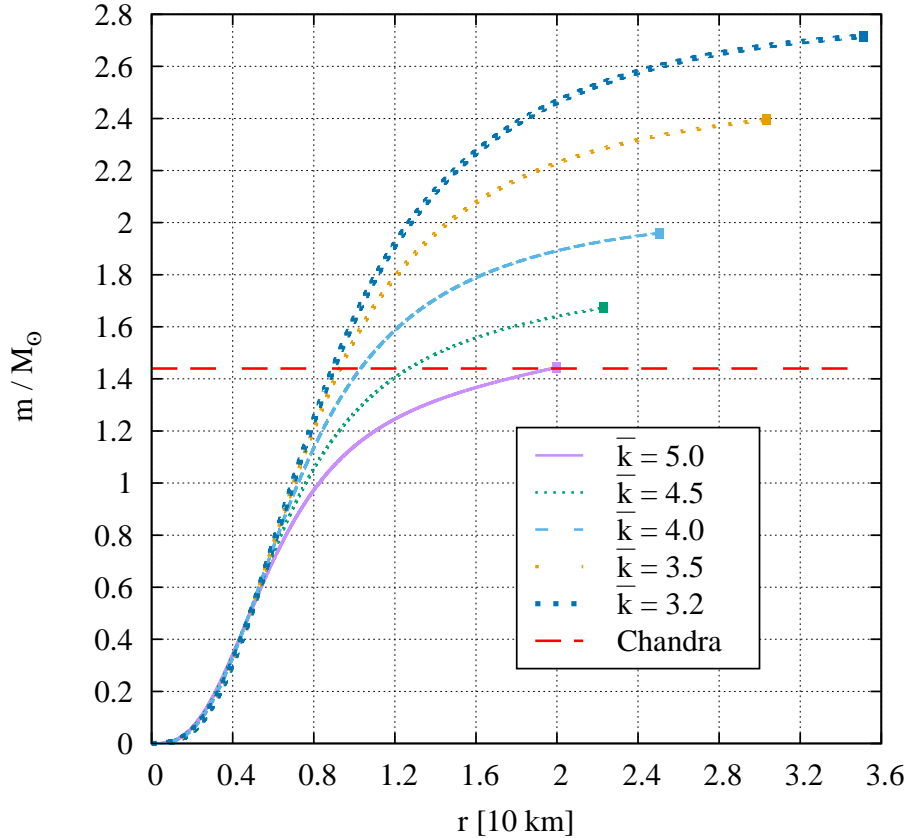}
  \caption{
  The mass $m$ as a function of radial coordinate $r$ with $\bar{\alpha}=0.01$ 
  and $\gamma\sim0.75$, where the value 1.44 is the Chandrasekhar (Chandra) limit (long-dashed line).} 
 \label{fig:III}
\end{figure}

%--------------------%
In FIG.~\ref{fig:I}, we illustrate the deviation of the interior region of the star 
in the $R^2$ model from GR. 
The profiles of the {scalar curvature} $\bar{R}$ and 
its derivative $\bar{R}'$ are shown in FIG.~\ref{fig:I}a.
Clearly, these two quantities satisfy the
boundary conditions $\bar{R}(x_s)=0$ and $\bar{R}'(x_s)=0$.
The results of the negative trace of the 
energy momentum tensor in GR and the $R^2$ model in the interior of the NSs
are displayed in FIG.~\ref{fig:I}b, illustrating similar behaviors.
However, the conduct of the scalar curvature 
in the $R^2$ model is different from that of 
GR with $R = -\,8\pi T = \kappa\,(\rho-3p)$
due to the $R^2$ term.

%--------------------%
\begin{table}
  \caption{The results of the mass $\bar{M}$, radius $x_s$ and 
   Ricci curvature  $\bar{R}(0)$ at the center
  for different values of the polytropic constant $\bar{k}$ 
  with $\bar{\alpha}=0.01$ and $\gamma\sim0.75$ in the $R^2$ model.}
 \begin{ruledtabular}
 \begin{tabular}{c c c c c c}
  %\hline\hline
  % inserts table 
  %heading
  $\bar{k}$   &  5.0    &  4.5    &  4.0    &  3.5    &  3.2 \\ [0.5ex] 
  \hline
  $\bar{M}$  &  1.444  &  1.673  &  1.959  &  2.394  &  2.716 \\
  $x_{s}$    &  1.999  &  2.228  &  2.504  &  3.032  &  3.513 \\ 
  $\bar{R}(0)$ &  8.95  &  8.40  &  7.54  &  6.15  &  4.93 \\ [1ex]
 %\hline\hline
 \end{tabular}
 \end{ruledtabular}
 \label{table:II}
\end{table}

%--------------------%
For the density $\bar{\rho}$ and mass function $m/M_{\odot}$ profiles of the star, 
we exhibit $\bar{k}=5.0$ with $\bar{\alpha}=0.01$ and $0.0005$ in FIG.~\ref{fig:II}.
We see that the deviation of the density in the $R^2$ model from GR is small in FIG.~\ref{fig:II}a, whereas that of the resultant mass is large in FIG.~\ref{fig:II}b.
The end-points of the curves in FIG.~\ref{fig:II}b correspond to 
the mass $\bar{M}$ and radius $x_s$ shown in TABLE \ref{table:I}.
In the $R^2$ model, the mass function in (\ref{E:m-prime equation 2})
is deviated from GR due to the geometric effect of the effective density $\rho_{\text{eff}}$.

%--------------------%
In TABLE \ref{table:I}, the mass of the NS exceeds the Chandrasekhar limit 
($1.44\,M_{\odot}$) of the white dwarf \cite{Chandrasekhar}. 
Note that 
if the collapsing process is supplied only by gravity,
 the Chandrasekhar limit could be considered as 
a lower bound of the mass for a star
whose ultimate destiny is a NS or black hole.

According to our analysis in the $R^2$ model, which
allows a lighter NS than that in GR as shown in FIG.~\ref{fig:II}b. 
Furthermore, from the upper limit 
$\alpha\lesssim 1.47722\times 10^{7}\, \text{m}^2$, 
we find that the \emph{minimal} mass of the NS is around 1.44 $M_{\odot}$
for $\gamma\sim0.75$ and $\bar{k}=5.0$
(see solid line in FIG.~\ref{fig:II}b).

%--------------------%
By fixing $\alpha$ equal to the critical value 
$\bar{\alpha}=0.01$, we can analyze the properties of the the NS
in the minimal mass condition.
The profiles of the mass function $m$ of the radial coordinate $r$ 
with $\gamma\sim0.75$, $\bar{k}=5.0$, 4.5, 4.0, 3.5 and $3.2$ are shown in FIG.~\ref{fig:III}, respectively.
In TABLE \ref{table:II}, we list the mass $\bar{M}$ and radius $x_s$ of the NSs 
and their corresponding Ricci curvature $\bar{R}(0)$ at the center.
From the table, we observe that the mass becomes \emph{larger} 
as $\bar{k}$ gets smaller, whereas $\bar{R}(0)$ becomes smaller. We note that the case of $\bar{k}=3.0$
due to $\bar{\rho}>3\bar{p}$ for ordinary matter inside the NS has been excluded in our discussion.
On the other hand, we expect that 
the mass of the NS is not smaller than the Chandrasekhar limit and
the value $\bar{k}$ can not be larger than 5.0.
The reasonable maximal value of $\bar{k}$ can be determined as 5.0.

\end{subsection}
\end{section}

%-----------------------------------------------------------------------%
% Sec. IV - Conclusions
%-----------------------------------------------------------------------%
\begin{section}{Conclusions}
\label{sec:Conclusions}
We have addressed the $R^2$ model on a compact star, especially on the NS
through the junction conditions. We have solved the 
\emph{mTOV equation} rather than the perturbation method in the literature.
In order to satisfy the junction conditions 
(Schwarzschild conditions), the central pressure $p(0)$ and 
Ricci scalar $R(0)$ should be well-selected. In $f(R)$ gravity, 
more specifically, the $R^2$ model, the parameters $k$ and 
$\gamma$ in the \emph{polytropic} EoS can be constrained by $p(0)$ and $R(0)$ due to the 
coupled structure equations. With the junction condition, in particular, 
we have shown that there exists the solution of EoS $\bar{\rho} = \bar{k}\, \bar{p}^{\,\gamma}$
with $\bar{k}\sim5.0$ and $\gamma\sim0.75$.

%--------------------%
For the upper limit $\alpha = 1.47722\times 10^{7}\, \text{m}^2$, 
we have obtained the minimal mass of the NS.
Under $\bar{\rho} = 5.0\, \bar{p}^{\, 0.75}$, 
the typical value of the NS mass is {around 1.44} $M_{\odot}$.
We have shown that $\bar{k}$ has the maximal value of $\bar{k}=5.0$. 
In our discussion, we have only considered the 
\emph{ghost-free} $f(R)$ theories ($\alpha>0$).
%{\color{blue} In FIG.~\ref{fig:I}(b), as $\alpha$ is around zero  .... }. 
%We would like to note that
One could have heavier NSs when taking a negative $\alpha$ into account 
under the polytrope assumption in Ref.~\cite{Orellana:2013gn}.
For $\alpha >0$, our result of the polytropic EoS is consistent with 
that in Ref.~\cite{Orellana:2013gn}.

%--------------------%
Finally, we remark that in our derivation, we have obtained the same coupled 
structure equations (\ref{E:Coupled ode2}) as those  in Ref.~\cite{Ganguly:2013taa}
after some proper arrangements.
However, by fine-tuning the EoS, we have solved (\ref{E:Coupled ode2})
with the junction conditions (\ref{E:Boundary conditions}) \emph{directly} 
rather than the \emph{indirect} method used in Ref.~\cite{Ganguly:2013taa}.
\end{section}

%-----------------------------------------------------------------------%
% Acknowledgments
%-----------------------------------------------------------------------%
\begin{acknowledgments}
We thank Professor Radouane Gannouji for communications and discussions.
This work was partially supported by National Center for 
Theoretical Sciences and MoST
 (MoST-104-2112-M-007-003-MY3 and MoST-104-2112-M-009-020-MY3).
%and National Tsing Hua University (104N2724E1), Taiwan, R.O.C.
%The work was supported in part by National Center for Theoretical Sciences,  National Science Council (NSC-101-2112-M-007-006-MY3) and National Tsing-Hua University (102N2725E1),  Taiwan, R.O.C.
\end{acknowledgments}

%-----------------------------------------------------------------------%
% Bibliography
%-----------------------------------------------------------------------%


\begin{thebibliography}{99}

%---------- Accelerated expansion - type Ia supernovae ----------%
%\cite{Riess:1998cb}
\bibitem{Riess:1998cb} 
  A.~G.~Riess {\it et al.} [Supernova Search Team Collaboration],
  %``Observational evidence from supernovae for an accelerating universe and a cosmological constant,''
  Astron.\ J.\  {\bf 116}, 1009 (1998).
  %doi:10.1086/300499
  %[astro-ph/9805201].
  %%CITATION = doi:10.1086/300499;%%
  %8559 citations counted in INSPIRE as of 20 Nov 2015

%\cite{Perlmutter:1998np}
\bibitem{Perlmutter:1998np} 
  S.~Perlmutter {\it et al.} [Supernova Cosmology Project Collaboration],
  %``Measurements of Omega and Lambda from 42 high redshift supernovae,''
  Astrophys.\ J.\  {\bf 517}, 565 (1999).
  %doi:10.1086/307221
  %[astro-ph/9812133].
  %%CITATION = doi:10.1086/307221;%%
  %8854 citations counted in INSPIRE as of 20 Nov 2015

%---------- large scale structure - SDSS ----------%
%\cite{Tegmark:2003ud}
\bibitem{Tegmark:2003ud} 
  M.~Tegmark {\it et al.} [SDSS Collaboration],
  %``Cosmological parameters from SDSS and WMAP,''
  Phys.\ Rev.\ D {\bf 69}, 103501 (2004).
  %doi:10.1103/PhysRevD.69.103501
  %[astro-ph/0310723].
  %%CITATION = doi:10.1103/PhysRevD.69.103501;%%
  %2429 citations counted in INSPIRE as of 20 Nov 2015

%\cite{Seljak:2004xh}
\bibitem{Seljak:2004xh} 
  U.~Seljak {\it et al.} [SDSS Collaboration],
  %``Cosmological parameter analysis including SDSS Ly-alpha forest and galaxy bias: Constraints on the primordial spectrum of fluctuations, neutrino mass, and dark energy,''
  Phys.\ Rev.\ D {\bf 71}, 103515 (2005).
  %doi:10.1103/PhysRevD.71.103515
  %[astro-ph/0407372].
  %%CITATION = doi:10.1103/PhysRevD.71.103515;%%
  %776 citations counted in INSPIRE as of 20 Nov 2015

%---------- Baryon acoustic oscillations ----------%
%\cite{Eisenstein:2005su}
\bibitem{Eisenstein:2005su} 
  D.~J.~Eisenstein {\it et al.} [SDSS Collaboration],
  %``Detection of the baryon acoustic peak in the large-scale correlation function of SDSS luminous red galaxies,''
  Astrophys.\ J.\  {\bf 633}, 560 (2005).
  %doi:10.1086/466512
  %[astro-ph/0501171].
  %%CITATION = doi:10.1086/466512;%%
  %2275 citations counted in INSPIRE as of 20 Nov 2015

%---------- Cosmic microwave background - WMAP ----------%
%\cite{Spergel:2003cb}
\bibitem{Spergel:2003cb} 
  D.~N.~Spergel {\it et al.} [WMAP Collaboration],
  %``First year Wilkinson Microwave Anisotropy Probe (WMAP) observations: Determination of cosmological parameters,''
  Astrophys.\ J.\ Suppl.\  {\bf 148}, 175 (2003).
  %doi:10.1086/377226
  %[astro-ph/0302209].
  %%CITATION = doi:10.1086/377226;%%
  %7724 citations counted in INSPIRE as of 20 Nov 2015

%\cite{Spergel:2006hy}
\bibitem{Spergel:2006hy} 
  D.~N.~Spergel {\it et al.} [WMAP Collaboration],
  %``Wilkinson Microwave Anisotropy Probe (WMAP) three year results: implications for cosmology,''
  Astrophys.\ J.\ Suppl.\  {\bf 170}, 377 (2007).
  %doi:10.1086/513700
  %[astro-ph/0603449].
  %%CITATION = doi:10.1086/513700;%%
  %5904 citations counted in INSPIRE as of 20 Nov 2015

%\cite{Komatsu:2008hk}
\bibitem{Komatsu:2008hk} 
  E.~Komatsu {\it et al.} [WMAP Collaboration],
  %``Five-Year Wilkinson Microwave Anisotropy Probe (WMAP) Observations: Cosmological Interpretation,''
  Astrophys.\ J.\ Suppl.\  {\bf 180}, 330 (2009).
  %doi:10.1088/0067-0049/180/2/330
  %[arXiv:0803.0547 [astro-ph]].
  %%CITATION = doi:10.1088/0067-0049/180/2/330;%%
  %4173 citations counted in INSPIRE as of 20 Nov 2015
%{\color{red}  
%---------- Review of Modified gravity theories ----------%
%\cite{Nojiri:2006ri}
\bibitem{Nojiri:2006ri} 
  S.~Nojiri and S.~D.~Odintsov,
  %``Introduction to modified gravity and gravitational alternative for dark energy,''
  eConf C {\bf 0602061}, 06 (2006)
  [Int.\ J.\ Geom.\ Meth.\ Mod.\ Phys.\  {\bf 4}, 115 (2007)].
  %doi:10.1142/S0219887807001928
  %[hep-th/0601213].
  %%CITATION = doi:10.1142/S0219887807001928;%%
  %1692 citations counted in INSPIRE as of 29 Aug 2017

%\cite{Nojiri:2017ncd}
\bibitem{Nojiri:2017ncd} 
  S.~Nojiri, S.~D.~Odintsov and V.~K.~Oikonomou,
  %``Modified Gravity Theories on a Nutshell: Inflation, Bounce and Late-time Evolution,''
  Phys.\ Rept.\  {\bf 692}, 1 (2017).
  %doi:10.1016/j.physrep.2017.06.001
  %[arXiv:1705.11098 [gr-qc]].
  %%CITATION = doi:10.1016/j.physrep.2017.06.001;%%
  %25 citations counted in INSPIRE as of 29 Aug 2017
 %} 
%---------- Inflation ----------%
%\cite{Starobinsky:1980te}
\bibitem{Starobinsky:1980te} 
  A.~A.~Starobinsky,
  %``A New Type of Isotropic Cosmological Models Without Singularity,''
  Phys.\ Lett.\ B {\bf 91}, 99 (1980).
  %doi:10.1016/0370-2693(80)90670-X
  %%CITATION = doi:10.1016/0370-2693(80)90670-X;%%
  %2508 citations counted in INSPIRE as of 17 Nov 2015

%\cite{Guth:1980zm}
\bibitem{Guth:1980zm} 
  A.~H.~Guth,
  %``The Inflationary Universe: A Possible Solution to the Horizon and Flatness Problems,''
  Phys.\ Rev.\ D {\bf 23}, 347 (1981).
  %doi:10.1103/PhysRevD.23.347
  %%CITATION = doi:10.1103/PhysRevD.23.347;%%
  %5522 citations counted in INSPIRE as of 17 Nov 2015

%\cite{Linde:1981mu}
\bibitem{Linde:1981mu} 
  A.~D.~Linde,
  %``A New Inflationary Universe Scenario: A Possible Solution of the Horizon, Flatness, Homogeneity, Isotropy and Primordial Monopole Problems,''
  Phys.\ Lett.\ B {\bf 108}, 389 (1982).
  %doi:10.1016/0370-2693(82)91219-9
  %%CITATION = doi:10.1016/0370-2693(82)91219-9;%%
  %3368 citations counted in INSPIRE as of 17 Nov 2015

%\cite{Albrecht:1982wi}
\bibitem{Albrecht:1982wi} 
  A.~Albrecht and P.~J.~Steinhardt,
  %``Cosmology for Grand Unified Theories with Radiatively Induced Symmetry Breaking,''
  Phys.\ Rev.\ Lett.\  {\bf 48}, 1220 (1982).
  %doi:10.1103/PhysRevLett.48.1220
  %%CITATION = doi:10.1103/PhysRevLett.48.1220;%%
  %3034 citations counted in INSPIRE as of 17 Nov 2015

%========== R-squared ==========%
%\cite{Starobinsky:1987zz}
\bibitem{Starobinsky:1987zz} 
  A.~A.~Starobinsky and H.-J.~Schmidt,
  %``On a general vacuum solution of fourth-order gravity,''
  Class.\ Quant.\ Grav.\  {\bf 4}, 695 (1987).
  %%CITATION = CQGRD,4,695;%%
  %65 citations counted in INSPIRE as of 11 Nov 2015

%---------- Cosmological Constant and dark energy ----------%
%\cite{Weinberg:1988cp}
\bibitem{Weinberg:1988cp} 
  S.~Weinberg,
  %``The Cosmological Constant Problem,''
  Rev.\ Mod.\ Phys.\  {\bf 61}, 1 (1989).
  %doi:10.1103/RevModPhys.61.1
  %%CITATION = doi:10.1103/RevModPhys.61.1;%%
  %3017 citations counted in INSPIRE as of 20 Nov 2015

%\cite{Sahni:1999gb}
\bibitem{Sahni:1999gb} 
  V.~Sahni and A.~A.~Starobinsky,
  %``The Case for a positive cosmological Lambda term,''
  Int.\ J.\ Mod.\ Phys.\ D {\bf 9}, 373 (2000).
  %[astro-ph/9904398].
  %%CITATION = ASTRO-PH/9904398;%%
  %1519 citations counted in INSPIRE as of 20 Nov 2015

  %\cite{Carroll:2000fy}
\bibitem{Carroll:2000fy} 
  S.~M.~Carroll,
  %``The Cosmological constant,''
  Living Rev.\ Rel.\  {\bf 4}, 1 (2001).
  %doi:10.12942/lrr-2001-1
  %[astro-ph/0004075].
  %%CITATION = doi:10.12942/lrr-2001-1;%%
  %1072 citations counted in INSPIRE as of 20 Nov 2015

%\cite{Peebles:2002gy}
\bibitem{Peebles:2002gy} 
  P.~J.~E.~Peebles and B.~Ratra,
  %``The Cosmological constant and dark energy,''
  Rev.\ Mod.\ Phys.\  {\bf 75}, 559 (2003).
  %doi:10.1103/RevModPhys.75.559
  %[astro-ph/0207347].
  %%CITATION = doi:10.1103/RevModPhys.75.559;%%
  %2687 citations counted in INSPIRE as of 20 Nov 2015

%\cite{Padmanabhan:2002ji}
\bibitem{Padmanabhan:2002ji} 
  T.~Padmanabhan,
  %``Cosmological constant: The Weight of the vacuum,''
  Phys.\ Rept.\  {\bf 380}, 235 (2003).
  %doi:10.1016/S0370-1573(03)00120-0
  %[hep-th/0212290].
  %%CITATION = doi:10.1016/S0370-1573(03)00120-0;%%
  %1909 citations counted in INSPIRE as of 20 Nov 2015

%\cite{Copeland:2006wr}
\bibitem{Copeland:2006wr} 
  E.~J.~Copeland, M.~Sami and S.~Tsujikawa,
  %``Dynamics of dark energy,''
  Int.\ J.\ Mod.\ Phys.\ D {\bf 15}, 1753 (2006).
  %doi:10.1142/S021827180600942X
  %[hep-th/0603057].
  %%CITATION = doi:10.1142/S021827180600942X;%%
  %2736 citations counted in INSPIRE as of 20 Nov 2015

%---------- f(R) theories review ----------%
%\cite{Barrow:1983rx}
\bibitem{Barrow:1983rx} 
  J.~D.~Barrow and A.~C.~Ottewill,
  %``The Stability of General Relativistic Cosmological Theory,''
  J.\ Phys.\ A {\bf 16}, 2757 (1983).
  %%CITATION = JPAGA,A16,2757;%%
  %264 citations counted in INSPIRE as of 10 Nov 2015

%\cite{Sotiriou:2008rp}
\bibitem{Sotiriou:2008rp}
 T.~P.~Sotiriou and V.~Faraoni,
 %``f(R) Theories Of Gravity,'' 
 Rev.\ Mod.\ Phys.\  {\bf 82}, 451 (2010).
 %[arXiv:0805.1726 [gr-qc]]. 
 %%CITATION = ARXIV:0805.1726;%%

%\cite{DeFelice:2010aj}
\bibitem{DeFelice:2010aj}
  A.~De Felice and S.~Tsujikawa,
  %``f(R) theories,''
  Living Rev.\ Rel.\  {\bf 13}, 3 (2010).
  %[arXiv:1002.4928 [gr-qc]].
  %%CITATION = 00222,13,3;%%
  
%{\color{red}
%\cite{Nojiri:2010wj}
\bibitem{Nojiri:2010wj} 
  S.~Nojiri and S.~D.~Odintsov,
  %``Unified cosmic history in modified gravity: from F(R) theory to Lorentz non-invariant models,''
  Phys.\ Rept.\  {\bf 505}, 59 (2011).
  %doi:10.1016/j.physrep.2011.04.001
  %[arXiv:1011.0544 [gr-qc]].
  %%CITATION = doi:10.1016/j.physrep.2011.04.001;%%
  %1332 citations counted in INSPIRE as of 23 Feb 2017
%}

%---------- viable f(R) model ----------%
%========== Hu-Sawicki ==========%
%\cite{Hu:2007nk}
\bibitem{Hu:2007nk}
  W.~Hu and I.~Sawicki,
  %``Models of f(R) Cosmic Acceleration that Evade Solar-System Tests,''
  Phys.\ Rev.\  D {\bf 76}, 064004 (2007). 
  %arXiv:0705.1158 [astro-ph]].
  %%CITATION = PHRVA,D76,064004;%%

%========== Starobinsky ==========%
  %\cite{Starobinsky:2007hu}
\bibitem{Starobinsky:2007hu} 
  A.~A.~Starobinsky,
  %``Disappearing cosmological constant in f(R) gravity,''
  JETP Lett.\  {\bf 86}, 157 (2007).
  %doi:10.1134/S0021364007150027
  %[arXiv:0706.2041 [astro-ph]].
  %%CITATION = doi:10.1134/S0021364007150027;%%
  %535 citations counted in INSPIRE as of 20 Nov 2015

%{\color{red}  
  %========== Nojiri ==========%  
%\cite{Nojiri:2007as}
\bibitem{Nojiri:2007as} 
  S.~Nojiri and S.~D.~Odintsov,
  %``Unifying inflation with LambdaCDM epoch in modified f(R) gravity consistent with Solar System tests,''
  Phys.\ Lett.\ B {\bf 657}, 238 (2007).
  %doi:10.1016/j.physletb.2007.10.027
  %[arXiv:0707.1941 [hep-th]].
  %%CITATION = doi:10.1016/j.physletb.2007.10.027;%%
  %274 citations counted in INSPIRE as of 23 Feb 2017
%} 
%========== Tsujikawa ==========%
%\cite{Tsujikawa:2007xu}
\bibitem{Tsujikawa:2007xu} 
  S.~Tsujikawa,
  %``Observational signatures of f(R) dark energy models that satisfy cosmological and local gravity constraints,''
  Phys.\ Rev.\ D {\bf 77}, 023507 (2008).
  %doi:10.1103/PhysRevD.77.023507
  %[arXiv:0709.1391 [astro-ph]].
  %%CITATION = doi:10.1103/PhysRevD.77.023507;%%
  %232 citations counted in INSPIRE as of 16 Nov 2015
  
%{\color{red}
%========== Cognola ==========% 
%\cite{Cognola:2007zu}
\bibitem{Cognola:2007zu} 
  G.~Cognola, E.~Elizalde, S.~Nojiri, S.~D.~Odintsov, L.~Sebastiani and S.~Zerbini,
  %``A Class of viable modified f(R) gravities describing inflation and the onset of accelerated expansion,''
  Phys.\ Rev.\ D {\bf 77}, 046009 (2008).
  %doi:10.1103/PhysRevD.77.046009
  %[arXiv:0712.4017 [hep-th]].
  %%CITATION = doi:10.1103/PhysRevD.77.046009;%%
  %357 citations counted in INSPIRE as of 23 Feb 2017
 %}
%========== Exponential Gravity ==========%
%\cite{ExpGra}
\bibitem{ExpGra}
  %\cite{Linder:2009jz}
%\bibitem{Linder:2009jz}
  E.~V.~Linder,
  %``Exponential Gravity,''
  Phys.\ Rev.\  D {\bf 80}, 123528 (2009);\ 
%\textit{ibid}.\  {\bf 80}, 123528 (2009);\ 
%  [arXiv:0905.2962 [astro-ph.CO]].
  %%CITATION = PHRVA,D80,123528;%%
  %\cite{Yang:2010xq}
%\bibitem{Yang:2010xq}
  L.~Yang, C.~C.~Lee, L.~W.~Luo and C.~Q.~Geng,
  Phys.\ Rev.\  D {\bf 82}, 103515 (2010);\  
  %``Observational Constraints on Exponential Gravity,''
%  arXiv:1010.2058 [astro-ph.CO].
  %%CITATION = ARXIV:1010.2058;%%
%\cite{Bamba:2010zz}
%\bibitem{Bamba:2010zz} 
  K.~Bamba, C.~Q.~Geng and C.~C.~Lee,
  %``Phantom crossing in viable $f(R)$ theories,''
  Int.\ J.\ Mod.\ Phys.\ D {\bf 20}, 1339 (2011);\
  %doi:10.1142/S0218271811019517
  %[arXiv:1108.2557 [gr-qc]].
  %%CITATION = doi:10.1142/S0218271811019517;%%
  %10 citations counted in INSPIRE as of 13 Dec 2015
%\cite{Chen:2014tdy}
%\bibitem{Chen:2014tdy} 
  Y.~Chen, C.~Q.~Geng, C.~C.~Lee, L.~W.~Luo and Z.~H.~Zhu,
  %``Constraints on the exponential $f(R)$ model from latest Hubble parameter measurements,''
  Phys.\ Rev.\ D {\bf 91}, no. 4, 044019 (2015).
%  [arXiv:1407.4303 [astro-ph.CO]].
  %%CITATION = ARXIV:1407.4303;%%

%===== Polytropic EoS =====% 
%\cite{Lattimer:2012nd}
\bibitem{Lattimer:2012nd} 
  J.~M.~Lattimer,
  %``The nuclear equation of state and neutron star masses,''
  Ann.\ Rev.\ Nucl.\ Part.\ Sci.\  {\bf 62}, 485 (2012).
  %doi:10.1146/annurev-nucl-102711-095018
  %[arXiv:1305.3510 [nucl-th]].
  %%CITATION = doi:10.1146/annurev-nucl-102711-095018;%%
  %273 citations counted in INSPIRE as of 08 Dec 2016
  
  %\cite{Chen:2015zpa}
\bibitem{Chen:2015zpa} 
  W.~C.~Chen and J.~Piekarewicz,
  %``Compactness of Neutron Stars,''
  Phys.\ Rev.\ Lett.\  {\bf 115}, no. 16, 161101 (2015)
  %doi:10.1103/PhysRevLett.115.161101
  %[arXiv:1505.07436 [nucl-th]].
  %%CITATION = doi:10.1103/PhysRevLett.115.161101;%%
  %7 citations counted in INSPIRE as of 12 Dec 2016
  
  %\cite{Reisenegger:2015crq}
\bibitem{Reisenegger:2015crq} 
  A.~Reisenegger and F.~S.~Zepeda,
  %``Order-of-magnitude physics of neutron stars,''
  Eur.\ Phys.\ J.\ A {\bf 52}, no. 3, 52 (2016)
  %doi:10.1140/epja/i2016-16052-y
  %[arXiv:1511.08813 [astro-ph.SR]].
  %%CITATION = doi:10.1140/epja/i2016-16052-y;%%
  
  %\cite{Gandolfi:2015jma}
\bibitem{Gandolfi:2015jma} 
  S.~Gandolfi, A.~Gezerlis and J.~Carlson,
  %``Neutron Matter from Low to High Density,''
  Ann.\ Rev.\ Nucl.\ Part.\ Sci.\  {\bf 65}, 303 (2015)
  %doi:10.1146/annurev-nucl-102014-021957
  %[arXiv:1501.05675 [nucl-th]].
  %%CITATION = doi:10.1146/annurev-nucl-102014-021957;%%
  %22 citations counted in INSPIRE as of 12 Dec 2016
  
  %\cite{Hebeler:2010jx}
\bibitem{Hebeler:2010jx} 
  K.~Hebeler, J.~M.~Lattimer, C.~J.~Pethick and A.~Schwenk,
  %``Constraints on neutron star radii based on chiral effective field theory interactions,''
  Phys.\ Rev.\ Lett.\  {\bf 105}, 161102 (2010)
  %doi:10.1103/PhysRevLett.105.161102
  %[arXiv:1007.1746 [nucl-th]].
  %%CITATION = doi:10.1103/PhysRevLett.105.161102;%%
  %185 citations counted in INSPIRE as of 12 Dec 2016
  
  %\cite{Barausse:2007pn}
\bibitem{Barausse:2007pn} 
  E.~Barausse, T.~P.~Sotiriou and J.~C.~Miller,
  %``A No-go theorem for polytropic spheres in Palatini f(R) gravity,''
  Class.\ Quant.\ Grav.\  {\bf 25}, 062001 (2008)
  %doi:10.1088/0264-9381/25/6/062001
  %[gr-qc/0703132 [GR-QC]].
  %%CITATION = doi:10.1088/0264-9381/25/6/062001;%%
  %103 citations counted in INSPIRE as of 12 Dec 2016
  
  %\cite{Santos:2009nu}
\bibitem{Santos:2009nu} 
  E.~Santos,
  %``Quantum vacuum effects as generalized f(R) gravity. Application to stars,''
  Phys.\ Rev.\ D {\bf 81}, 064030 (2010).
  %doi:10.1103/PhysRevD.81.064030
  %[arXiv:0909.0120 [gr-qc]].
  %%CITATION = doi:10.1103/PhysRevD.81.064030;%%
  %17 citations counted in INSPIRE as of 20 Nov 2015

%===== Neutron star =====%
%\cite{Cooney:2009rr}
\bibitem{Cooney:2009rr} 
  A.~Cooney, S.~DeDeo and D.~Psaltis,
  %``Neutron Stars in f(R) Gravity with Perturbative Constraints,''
  Phys.\ Rev.\ D {\bf 82}, 064033 (2010).
  %doi:10.1103/PhysRevD.82.064033
  %[arXiv:0910.5480 [astro-ph.HE]].
  %%CITATION = doi:10.1103/PhysRevD.82.064033;%%
  %39 citations counted in INSPIRE as of 20 Nov 2015

%===== Relativistic star =====%
%\cite{Babichev:2009fi}
\bibitem{Babichev:2009fi} 
  E.~Babichev and D.~Langlois,
  %``Relativistic stars in f(R) and scalar-tensor theories,''
  Phys.\ Rev.\ D {\bf 81}, 124051 (2010).
  %doi:10.1103/PhysRevD.81.124051
  %[arXiv:0911.1297 [gr-qc]].
  %%CITATION = doi:10.1103/PhysRevD.81.124051;%%
  %53 citations counted in INSPIRE as of 20 Nov 2015

%===== Neutron star =====%
%\cite{Reijonen:2009hi}
\bibitem{Reijonen:2009hi} 
  V.~Reijonen,
  %``On white dwarfs and neutron stars in Palatini f(R) gravity,''
  arXiv:0912.0825 [gr-qc].
  %%CITATION = ARXIV:0912.0825;%%
  %13 citations counted in INSPIRE as of 20 Nov 2015
  
  %===== Neutron star in Starobinsky model =====%
%\cite{Orellana:2013gn}
\bibitem{Orellana:2013gn} 
  M.~Orellana, F.~Garcia, F.~A.~Teppa Pannia and G.~E.~Romero,
  %``Structure of neutron stars in $R$-squared gravity,''
  Gen.\ Rel.\ Grav.\  {\bf 45}, 771 (2013).
  %doi:10.1007/s10714-013-1501-5
  %[arXiv:1301.5189 [astro-ph.CO]].
  %%CITATION = doi:10.1007/s10714-013-1501-5;%%
  %17 citations counted in INSPIRE as of 20 Nov 2015
  
  %===== Relativistic star =====%
%\cite{Alavirad:2013paa}
\bibitem{Alavirad:2013paa} 
  H.~Alavirad and J.~M.~Weller,
  %``Modified gravity with logarithmic curvature corrections and the structure of relativistic stars,''
  Phys.\ Rev.\ D {\bf 88}, no. 12, 124034 (2013).
  %doi:10.1103/PhysRevD.88.124034
  %[arXiv:1307.7977].
  %%CITATION = doi:10.1103/PhysRevD.88.124034;%%
  %19 citations counted in INSPIRE as of 20 Nov 2015
  
  %===== Neutron star in Starobinsky model =====%
%\cite{Ganguly:2013taa}
\bibitem{Ganguly:2013taa} 
  A.~Ganguly, R.~Gannouji, R.~Goswami and S.~Ray,
  %``Neutron stars in the Starobinsky model,''
  Phys.\ Rev.\ D {\bf 89}, no. 6, 064019 (2014).
  %doi:10.1103/PhysRevD.89.064019
  %[arXiv:1309.3279 [gr-qc]].
  %%CITATION = doi:10.1103/PhysRevD.89.064019;%%
  %9 citations counted in INSPIRE as of 20 Nov 2015
  
  %===== Neutron star in Starobinsky model =====%
%\cite{Yazadjiev:2014cza}
\bibitem{Yazadjiev:2014cza} 
  S.~S.~Yazadjiev, D.~D.~Doneva, K.~D.~Kokkotas and K.~V.~Staykov,
  %``Non-perturbative and self-consistent models of neutron stars in R-squared gravity,''
  JCAP {\bf 1406}, 003 (2014).
  %doi:10.1088/1475-7516/2014/06/003
  %[arXiv:1402.4469 [gr-qc]].
  %%CITATION = doi:10.1088/1475-7516/2014/06/003;%%
  %21 citations counted in INSPIRE as of 20 Nov 2015
  
  %\cite{Hebeler:2013nza}
\bibitem{Hebeler:2013nza} 
  K.~Hebeler, J.~M.~Lattimer, C.~J.~Pethick and A.~Schwenk,
  %``Equation of state and neutron star properties constrained by nuclear physics and observation,''
  Astrophys.\ J.\  {\bf 773}, 11 (2013)
  %doi:10.1088/0004-637X/773/1/11
  %[arXiv:1303.4662 [astro-ph.SR]].
  %%CITATION = doi:10.1088/0004-637X/773/1/11;%%
  %135 citations counted in INSPIRE as of 11 Dec 2016

%---------- White dwarf, neutron star and compact star ----------%
%\cite{Chandrasekhar}
\bibitem{Chandrasekhar} 
%\cite{Chandrasekhar:1931ih}
%\bibitem{Chandrasekhar:1931ih} 
  S.~Chandrasekhar,
  %``The maximum mass of ideal white dwarfs,''
  Astrophys.\ J.\  {\bf 74}, 81 (1931);
  %doi:10.1086/143324
  %%CITATION = doi:10.1086/143324;%%
  %181 citations counted in INSPIRE as of 20 Nov 2015
%\cite{Chandrasekhar:1935zz}
%\bibitem{Chandrasekhar:1935zz} 
  %S.~Chandrasekhar,
  %``The highly collapsed configurations of a stellar mass (Second paper),''
  Mon.\ Not.\ Roy.\ Astron.\ Soc.\  {\bf 95}, 207 (1935).
  %%CITATION = MNRAA,95,207;%%
  %84 citations counted in INSPIRE as of 17 Oct 2015

%\cite{Oppenheimer:1939ne}
\bibitem{Oppenheimer:1939ne} 
  J.~R.~Oppenheimer and G.~M.~Volkoff,
  %``On Massive neutron cores,''
  Phys.\ Rev.\  {\bf 55}, 374 (1939).
  %doi:10.1103/PhysRev.55.374
  %%CITATION = doi:10.1103/PhysRev.55.374;%%
  %912 citations counted in INSPIRE as of 20 Nov 2015

%---------- Stars in modified gravity ----------%

%===== Singular problem =====%
%{\color{red}
%\cite{Briscese:2006xu}
\bibitem{Briscese:2006xu} 
  F.~Briscese, E.~Elizalde, S.~Nojiri and S.~D.~Odintsov,
  %``Phantom scalar dark energy as modified gravity: Understanding the origin of the Big Rip singularity,''
  Phys.\ Lett.\ B {\bf 646}, 105 (2007).
  %doi:10.1016/j.physletb.2007.01.013
  %[hep-th/0612220].
  %%CITATION = doi:10.1016/j.physletb.2007.01.013;%%
  %179 citations counted in INSPIRE as of 23 Feb 2017
  
%\cite{Nojiri:2008fk}
\bibitem{Nojiri:2008fk} 
  S.~Nojiri and S.~D.~Odintsov,
  %``The Future evolution and finite-time singularities in F(R)-gravity unifying the inflation and cosmic acceleration,''
  Phys.\ Rev.\ D {\bf 78}, 046006 (2008).
  %doi:10.1103/PhysRevD.78.046006
  %[arXiv:0804.3519 [hep-th]].
  %%CITATION = doi:10.1103/PhysRevD.78.046006;%%
  %181 citations counted in INSPIRE as of 23 Feb 2017
% } 
%\cite{Frolov:2008uf}
\bibitem{Frolov:2008uf} 
  A.~V.~Frolov,
  %``A Singularity Problem with f(R) Dark Energy,''
  Phys.\ Rev.\ Lett.\  {\bf 101}, 061103 (2008).
  %doi:10.1103/PhysRevLett.101.061103
  %[arXiv:0803.2500 [astro-ph]].
  %%CITATION = doi:10.1103/PhysRevLett.101.061103;%%
  %119 citations counted in INSPIRE as of 20 Nov 2015
  

  %===== Relativistic star =====%
%\cite{Kobayashi:2008tq}
\bibitem{Kobayashi:2008tq} 
T.~Kobayashi and K.~i.~Maeda,
%"Relativistic stars in f(R) gravity, and absence thereof,''
Phys.\ Rev.\ D {\bf 78}, 064019 (2008).
%arXiv:0807.2503 [astro-ph].

%===== Relativistic star =====%
%\cite{Babichev:2009td}
\bibitem{Babichev:2009td} 
  E.~Babichev and D.~Langlois,
  %``Relativistic stars in f(R) gravity,''
  Phys.\ Rev.\ D {\bf 80}, 121501 (2009)
  [Phys.\ Rev.\ D {\bf 81}, 069901 (2010)].
  %doi:10.1103/PhysRevD.81.069901, 10.1103/PhysRevD.80.121501
  %[arXiv:0904.1382 [gr-qc]].
  %%CITATION = doi:10.1103/PhysRevD.81.069901, 10.1103/PhysRevD.80.121501;%%
  %55 citations counted in INSPIRE as of 20 Nov 2015
  

%===== Relativistic star =====%
%\cite{Upadhye:2009kt}
\bibitem{Upadhye:2009kt} 
  A.~Upadhye and W.~Hu,
  %``The existence of relativistic stars in f(R) gravity,''
  Phys.\ Rev.\ D {\bf 80}, 064002 (2009).
  %doi:10.1103/PhysRevD.80.064002
  %[arXiv:0905.4055 [astro-ph.CO]].
  %%CITATION = doi:10.1103/PhysRevD.80.064002;%%
  %52 citations counted in INSPIRE as of 20 Nov 2015


%===== Neutron star =====%
%===== Constraint of f(R) =====%
%\cite{Arapoglu:2010rz}
\bibitem{Arapoglu:2010rz} 
  A.~S.~Arapoglu, C.~Deliduman and K.~Y.~Eksi,
  %``Constraints on Perturbative f(R) Gravity via Neutron Stars,''
  JCAP {\bf 1107}, 020 (2011).
  %doi:10.1088/1475-7516/2011/07/020
  %[arXiv:1003.3179 [gr-qc]].
  %%CITATION = doi:10.1088/1475-7516/2011/07/020;%%
  %33 citations counted in INSPIRE as of 20 Nov 2015

%===== Relativistic star =====%
%\cite{Jaime:2010kn}
\bibitem{Jaime:2010kn} 
  L.~G.~Jaime, L.~Patino and M.~Salgado,
  %``Robust approach to f(R) gravity,''
  Phys.\ Rev.\ D {\bf 83}, 024039 (2011)
  %doi:10.1103/PhysRevD.83.024039
  %[arXiv:1006.5747 [gr-qc]].
  %%CITATION = doi:10.1103/PhysRevD.83.024039;%%
  %22 citations counted in INSPIRE as of 23 Nov 2015
  
%===== Neutron star =====%
%\cite{Santos:2011ye}
\bibitem{Santos:2011ye} 
  E.~Santos,
  %``Neutron stars in generalized f(R) gravity,''
  Astrophys.\ Space Sci.\  {\bf 341}, 411 (2012).
  %doi:10.1007/s10509-012-1049-y
  %[arXiv:1104.2140 [gr-qc]].
  %%CITATION = doi:10.1007/s10509-012-1049-y;%%
  %15 citations counted in INSPIRE as of 20 Nov 2015

%===== Constraint of f(R) =====%
%\cite{Cheoun:2013tsa}
\bibitem{Cheoun:2013tsa} 
  M.~K.~Cheoun, C.~Deliduman, C.~Güngör, V.~Keleş, C.~Y.~Ryu, T.~Kajino and G.~J.~Mathews,
  %``Neutron stars in a perturbative $f(R)$ gravity model with strong magnetic fields,''
  JCAP {\bf 1310}, 021 (2013).
  %doi:10.1088/1475-7516/2013/10/021
  %[arXiv:1304.1871 [astro-ph.HE]].
  %%CITATION = doi:10.1088/1475-7516/2013/10/021;%%
  %16 citations counted in INSPIRE as of 20 Nov 2015

%===== Neutron star =====%
%\cite{Astashenok:2013vza}
\bibitem{Astashenok:2013vza} 
  A.~V.~Astashenok, S.~Capozziello and S.~D.~Odintsov,
  %``Further stable neutron star models from f(R) gravity,''
  JCAP {\bf 1312}, 040 (2013);
  %doi:10.1088/1475-7516/2013/12/040
  %[arXiv:1309.1978 [gr-qc]].
  %%CITATION = doi:10.1088/1475-7516/2013/12/040;%%
  %41 citations counted in INSPIRE as of 20 Nov 2015
%\cite{Astashenok:2014pua}
%\bibitem{Astashenok:2014pua} 
  %A.~V.~Astashenok, S.~Capozziello and S.~D.~Odintsov,
  %``Maximal neutron star mass and the resolution of the hyperon puzzle in modified gravity,''
  Phys.\ Rev.\ D {\bf 89}, no. 10, 103509 (2014);
  %doi:10.1103/PhysRevD.89.103509
  %[arXiv:1401.4546 [gr-qc]].
  %%CITATION = doi:10.1103/PhysRevD.89.103509;%%
  %31 citations counted in INSPIRE as of 20 Nov 2015
%\cite{Astashenok:2014gda}
%\bibitem{Astashenok:2014gda} 
  %A.~V.~Astashenok, S.~Capozziello and S.~D.~Odintsov,
  %``Magnetic Neutron Stars in f(R) gravity,''
  Astrophys.\ Space Sci.\  {\bf 355}, no. 2, 333 (2015).
  %doi:10.1007/s10509-014-2182-6
  %[arXiv:1405.6663 [gr-qc]].
  %%CITATION = doi:10.1007/s10509-014-2182-6;%%
  %10 citations counted in INSPIRE as of 20 Nov 2015

%===== Neutron star in Starobinsky model =====%
%{\color{red}
%\cite{Staykov:2014mwa}
\bibitem{Staykov:2014mwa} 
  K.~V.~Staykov, D.~D.~Doneva, S.~S.~Yazadjiev and K.~D.~Kokkotas,
  %``Slowly rotating neutron and strange stars in $R^2$ gravity,''
  JCAP {\bf 1410}, no. 10, 006 (2014).
  %doi:10.1088/1475-7516/2014/10/006
  %[arXiv:1407.2180 [gr-qc]].
  %%CITATION = doi:10.1088/1475-7516/2014/10/006;%%
  %24 citations counted in INSPIRE as of 23 Feb 2017
%}  
%\cite{Yazadjiev:2015zia}
\bibitem{Yazadjiev:2015zia} 
  S.~S.~Yazadjiev, D.~D.~Doneva and K.~D.~Kokkotas,
  %``Rapidly rotating neutron stars in R-squared gravity,''
  Phys.\ Rev.\ D {\bf 91}, no. 8, 084018 (2015).
  %doi:10.1103/PhysRevD.91.084018
  %[arXiv:1501.04591 [gr-qc]].
  %%CITATION = doi:10.1103/PhysRevD.91.084018;%%
  %6 citations counted in INSPIRE as of 20 Nov 2015
  
%===== Relativistic star =====%
%\cite{Goswami:2015dma}
\bibitem{Goswami:2015dma} 
  R.~Goswami, S.~D.~Maharaj and A.~M.~Nzioki,
  %``Buchdahl-Bondi limit in modified gravity: Packing extra effective mass in relativistic compact stars,''
  Phys.\ Rev.\ D {\bf 92}, 064002 (2015).
  %doi:10.1103/PhysRevD.92.064002
  %[arXiv:1506.04043 [gr-qc]].
  %%CITATION = doi:10.1103/PhysRevD.92.064002;%%
  %1 citations counted in INSPIRE as of 20 Nov 2015

%\cite{Staykov:2015mma}
\bibitem{Staykov:2015mma} 
    K.~Staykov, K.~Y.~Ekşi, S.~S.~Yazadjiev, M.~M.~Türkoğlu and A.~S.~Arapoğlu,
  %``Moment of inertia of neutron star crust in alternative and modified theories of gravity,''
  Phys.\ Rev.\ D {\bf 94}, no. 2, 024056 (2016)
  %doi:10.1103/PhysRevD.94.024056
  %[arXiv:1507.05878 [gr-qc]].
  %%CITATION = doi:10.1103/PhysRevD.94.024056;%%
  %1 citations counted in INSPIRE as of 12 Dec 2016

  %\cite{Capozziello:2015yza}
\bibitem{Capozziello:2015yza} 
  S.~Capozziello, M.~De Laurentis, R.~Farinelli and S.~D.~Odintsov,
  %``Mass-radius relation for neutron stars in f(R) gravity,''
  Phys.\ Rev.\ D {\bf 93}, no. 2, 023501 (2016)
  %doi:10.1103/PhysRevD.93.023501
  %[arXiv:1509.04163 [gr-qc]].
  %%CITATION = doi:10.1103/PhysRevD.93.023501;%%
  %10 citations counted in INSPIRE as of 12 Dec 2016
  
%{\color{red}
%\cite{Yazadjiev:2015xsj}
\bibitem{Yazadjiev:2015xsj} 
  S.~S.~Yazadjiev and D.~D.~Doneva,
  %``Comment on "The Mass-Radius relation for Neutron Stars in $f(R)$ gravity" by S. Capozziello, M. De Laurentis, R. Farinelli and S. Odintsov,''
  arXiv:1512.05711 [gr-qc].
  %%CITATION = ARXIV:1512.05711;%%
  %2 citations counted in INSPIRE as of 23 Feb 2017
 % }  
%{\color{red}
%\cite{Hendi:2015pua}
\bibitem{Hendi:2015pua} 
  S.~H.~Hendi, G.~H.~Bordbar, B.~Eslam Panah and M.~Najafi,
  %``Dilatonic Equation of Hydrostatic Equilibrium and Neutron Star Structure,''
  Astrophys.\ Space Sci.\  {\bf 358}, no. 2, 30 (2015).
  %doi:10.1007/s10509-015-2429-x
  %[arXiv:1503.01011 [gr-qc]].
  %%CITATION = doi:10.1007/s10509-015-2429-x;%%
  %6 citations counted in INSPIRE as of 23 Feb 2017
  
%\cite{Hendi:2015vta}
\bibitem{Hendi:2015vta} 
  S.~H.~Hendi, G.~H.~Bordbar, B.~E.~Panah and S.~Panahiyan,
  %``Modified TOV in gravity's rainbow: properties of neutron stars and dynamical stability conditions,''
  JCAP {\bf 1609}, no. 09, 013 (2016).
  %doi:10.1088/1475-7516/2016/09/013
  %[arXiv:1509.05145 [hep-th]].
  %%CITATION = doi:10.1088/1475-7516/2016/09/013;%%
  %10 citations counted in INSPIRE as of 23 Feb 2017

%\cite{Bordbar:2015wva}
\bibitem{Bordbar:2015wva} 
  G.~H.~Bordbar, S.~H.~Hendi and B.~Eslam Panah,
  %``Neutron stars in Einstein-$\Lambda$ gravity: the cosmological constant effects,''
  Eur.\ Phys.\ J.\ Plus {\bf 131}, no. 9, 315 (2016).
  %doi:10.1140/epjp/i2016-16315-0
  %[arXiv:1502.02929 [gr-qc]].
  %%CITATION = doi:10.1140/epjp/i2016-16315-0;%%
  %5 citations counted in INSPIRE as of 23 Feb 2017
  
%\cite{Hendi:2017ibm}
\bibitem{Hendi:2017ibm} 
  S.~H.~Hendi, G.~H.~Bordbar, B.~Eslam Panah and S.~Panahiyan,
  %``Neutron stars structure in the context of massive gravity,''
  arXiv:1701.01039 [gr-qc].
  %%CITATION = ARXIV:1701.01039;%%
  %1 citations counted in INSPIRE as of 23 Feb 2017
%}

%---------- Junction conditions for F(R)-gravity ----------%
%\cite{Senovilla:2013vra}
\bibitem{Senovilla:2013vra} 
  J.~M.~M.~Senovilla,
  %``Junction conditions for F(R)-gravity and their consequences,''
  Phys.\ Rev.\ D {\bf 88}, 064015 (2013).
  %doi:10.1103/PhysRevD.88.064015
  %[arXiv:1303.1408 [gr-qc]].
  %%CITATION = doi:10.1103/PhysRevD.88.064015;%%
  %9 citations counted in INSPIRE as of 20 Nov 2015
  
 %\cite{Deruelle:2007pt}
%{\color{red}
\bibitem{Deruelle:2007pt} 
  N.~Deruelle, M.~Sasaki and Y.~Sendouda,
  %``Junction conditions in f(R) theories of gravity,''
  Prog.\ Theor.\ Phys.\  {\bf 119}, 237 (2008).
  %}
  %doi:10.1143/PTP.119.237
  %[arXiv:0711.1150 [gr-qc]].
  %%CITATION = doi:10.1143/PTP.119.237;%%
  %54 citations counted in INSPIRE as of 23 May 2017
%{\color{red}  
%---------- Non-perturbative methods in f(R) ----------%
%\cite{Astashenok:2014dja}
\bibitem{Astashenok:2014dja} 
  A.~V.~Astashenok, S.~Capozziello and S.~D.~Odintsov,
  %``Nonperturbative models of quark stars in $f$(R) gravity,''
  Phys.\ Lett.\ B {\bf 742}, 160 (2015).
  %doi:10.1016/j.physletb.2015.01.030
  %[arXiv:1412.5453 [gr-qc]].
  %%CITATION = doi:10.1016/j.physletb.2015.01.030;%%
  %19 citations counted in INSPIRE as of 29 Aug 2017
  
%\cite{Capozziello:2011nr}
\bibitem{Capozziello:2011nr} 
  S.~Capozziello, M.~De Laurentis, S.~D.~Odintsov and A.~Stabile,
  %``Hydrostatic equilibrium and stellar structure in f(R)-gravity,''
  Phys.\ Rev.\ D {\bf 83}, 064004 (2011).
  %doi:10.1103/PhysRevD.83.064004
  %[arXiv:1101.0219 [gr-qc]].
  %%CITATION = doi:10.1103/PhysRevD.83.064004;%%
  %65 citations counted in INSPIRE as of 29 Aug 2017
  
%\cite{Astashenok:2017dpo}
\bibitem{Astashenok:2017dpo} 
  A.~V.~Astashenok, A.~de la Cruz-Dombriz and S.~D.~Odintsov,
  %``The realistic models of relativistic stars in f(R) = R + alpha R^2 gravity,''
  arXiv:1704.08311 [gr-qc].
  %%CITATION = ARXIV:1704.08311;%%
  %4 citations counted in INSPIRE as of 29 Aug 2017
%}
%===== Jebsen-Birkhoff theorem in f(R) =====%  
%\cite{Nzioki:2013lca}
\bibitem{Nzioki:2013lca} 
  A.~M.~Nzioki, R.~Goswami and P.~K.~S.~Dunsby,
  %``Jebsen-Birkhoff theorem and its stability in f(R) gravity,''
  Phys.\ Rev.\ D {\bf 89}, no. 6, 064050 (2014).
  %doi:10.1103/PhysRevD.89.064050
  %[arXiv:1312.6790 [gr-qc]].
  %%CITATION = doi:10.1103/PhysRevD.89.064050;%%
  %14 citations counted in INSPIRE as of 27 Jun 2017
  
%===== Constraint of f(R) =====%
%\cite{Naf:2010zy}
\bibitem{Naf:2010zy} 
  J.~Naf and P.~Jetzer,
  %``On the 1/c Expansion of f(R) Gravity,''
  Phys.\ Rev.\ D {\bf 81}, 104003 (2010);
  %doi:10.1103/PhysRevD.81.104003
  %[arXiv:1004.2014 [gr-qc]].
  %%CITATION = doi:10.1103/PhysRevD.81.104003;%%
  %33 citations counted in INSPIRE as of 11 Dec 2016
%\cite{Naf:2011za}
%\bibitem{Naf:2011za} 
  %J.~Naf and P.~Jetzer,
  %``On Gravitational Radiation in Quadratic $f(R)$ Gravity,''
  Phys.\ Rev.\ D {\bf 84}, 024027 (2011).
  %doi:10.1103/PhysRevD.84.024027
  %[arXiv:1104.2200 [gr-qc]].
  %%CITATION = doi:10.1103/PhysRevD.84.024027;%%
  %14 citations counted in INSPIRE as of 20 Nov 2015

%----------Some experiments & observations ----------%

%\cite{Everitt:2009}
\bibitem{Everitt:2009} 
  Everitt, C.W.F., Adams, M., Bencze, W. {\it et al.},
  %``Gravity Probe B Data Analysis,''
  Space Sci Rev (2009) 148: 53. 
  %doi:10.1007/s11214-009-9524-7

%\cite{Breton:2008xy}
\bibitem{Breton:2008xy} 
  R.~P.~Breton {\it et al.},
  %``Relativistic Spin Precession in the Double Pulsar,''
  Science {\bf 321}, 104 (2008)
  %doi:10.1126/science.1159295
  %[arXiv:0807.2644 [astro-ph]].
  %%CITATION = doi:10.1126/science.1159295;%%
  %68 citations counted in INSPIRE as of 11 Dec 2016


\end{thebibliography}
\end{document}